\documentclass[
 reprint,
amsmath,
amssymb,
aps,
pra,
]{revtex4-1}
\usepackage{comment}
\usepackage{graphicx}
\usepackage{dcolumn}
\usepackage{bm}
\usepackage{color}
\usepackage{multirow}
\usepackage{geometry}
\usepackage{multirow,array}
\usepackage{ulem}

\newcolumntype{P}[1]{>{\centering\arraybackslash}p{#1}}
\begin{document}


\title{Critical and Topological Phases of Dimerized Kitaev Chain in Presence of Quasiperiodic Potential}

\author{Shilpi Roy$^{1}$, Sk Noor Nabi$^{2}$, and  Saurabh Basu$^{1}$}
\affiliation{$^{1}$Department of Physics, Indian Institute of Technology Guwahati-Guwahati, 781039 Assam, India}
\affiliation{$^{2}$Department of Physics, Indian Institute of Technology Kharagpur, Kharagpur - 721302, West Bengal, India}

\date{\today}

\begin{abstract}
We investigate localization and topological properties of a dimerized Kitaev chain with $p$-wave superconducting correlations and a quasiperiodically modulated chemical potential. With regard to the localization studies, we demonstrate the existence of distinct phases, such as, the extended phase, the critical (intermediate) phase, and the localized phase that arise due to the competition between the dimerization and the onsite quasiperiodic potential. Most interestingly, the critical phase comprises of two different mobility edges that are found to exist between the extended to the localized phase, and between the critical (multifractal) and localized phases. We perform our analysis employing the inverse and the normalized participation ratios, fractal dimension, and the level spacing. Subsequently, a finite-size analysis is done to provide support of our findings. Furthermore, we study the topological properties of the zero-energy edge modes via computing the real-space winding number and number of the Majorana zero modes present in the system. We specifically illustrate that our model exhibits a phase transition from a topologically trivial to a non-trivial phase (topological Anderson phase) beyond a critical dimerization strength under the influence of the quasiperiodic potential strength. Finally, in presence of a large potential, we demonstrate that the system undergoes yet another transition from the topologically non-trivial to an Anderson localized phase. Thus, we believe that our results will aid exploration of fundamentally different physics pertaining to the critical and the topological Anderson phases.
\end{abstract}

\maketitle


\section{\label{sec:level1} Introduction}

Quasiperiodic (QP) potential lies in between the completely periodic and the random potentials regime \cite{SOKOLOFF1985189}.
While a periodic potential entails Bloch states, in contrast the random potential induces a complete localization of all the single particle eigenstates in one dimension (1D) and two dimensions (2D). The latter is known as the Anderson localization \cite{PhysRev.109.1492}. In three dimensions (3D), a phase transition from an extended to a localized phase is possible, resulting in the emergence of a mobility edge \cite{PhysRevLett.42.673, mott1987mobility}. The mobility edge denotes a critical point for the onset of localization transitions, that separate the extended states from the localized ones.

Due to the experimental accessibility, the QP potential arises in vast range of fields, such as, in optical \cite{roati2008anderson,PhysRevLett.120.160404,Modugno_2009,PhysRevA.72.053607,viebahn2019matter,yao2019critical,PhysRevLett.122.170403}, photonic\cite{PhysRevLett.103.013901,PhysRevLett.110.076403,PhysRevB.91.064201,PhysRevLett.109.106402,wang2020localization}, phononic\cite{PhysRevLett.125.224301,PhysRevLett.122.095501}, cavity-polariton \cite{PhysRevLett.112.146404,goblot2020emergence} cases, and more recently found in moir\'e lattices \cite{balents2020superconductivity}. The transport properties of the system in the backdrop of this QP potential are largely studied via the Aubry-Andr\'e (AA) model \cite{aubry1980analyticity}. This tight-binding model comprises of a nearest-neighbor hopping in the presence of the QP potential. Interestingly, a phase transition happens from a completely extended phase to a totally localized phase on a one-dimensional chain at a critical potential strength. Thus, the model shows no mobility edge in one dimension. However, An energy-dependent mobility edge appears as one goes beyond this approximation as has been shown in several theoretical models \cite{PhysRevA.80.021603,PhysRevLett.104.070601,PhysRevB.83.075105,PhysRevLett.114.146601,PhysRevB.96.085119,PhysRevB.101.064203,wang2020one,PhysRevB.103.L060201,PhysRevB.91.235134} and experimental situations \cite{an2018engineering,PhysRevLett.126.040603,PhysRevLett.122.170403,PhysRevLett.120.160404}. In addition, various generalizations of the model have also been studied to understand the localization transition in quasiperiodic systems \cite{PhysRevA.72.053607,deng2019one,szabo2018non,10.21468/SciPostPhys.4.5.025,PhysRevB.101.174203,PhysRevB.96.180204,PhysRevLett.126.106803,roy2022critical, PhysRevB.105.L220201}. Hence, a metal-insulator transition albeit being prohibited in presence of a 1D random potential is possible for QP potentials.

Other than the existence of the two emerging phases such as, the extended and the localized phases, QP potential is also known to host another intriguing phase, namely, the critical (intermediate) phase \cite{PhysRevLett.126.106803,roy2022critical}. A critical phase is characterized by the coexistence of different phases, either with a precise boundary and mobility edges, or without a boundary resulting in the formation of a mixed phase \cite{PhysRevB.91.014108,PhysRevLett.104.070601,wang2020one,PhysRevB.101.014205}. It is known that the eigenstates at the mobility edge are multifractal in nature \cite{RevModPhys.80.1355}. Interestingly, a QP potential can host regimes comprising of critical (multifractal) states in a wide range of parameter space, leading to a multifractal phase \cite{PhysRevB.91.014108,10.21468/SciPostPhys.12.1.027,deng2019one,PhysRevLett.110.146404,PhysRevLett.125.073204}. These multifractal states are fundamentally different from the extended and the localized states, thereby making it feasible to explore new opportunities in different branches of physics, such as, non-ergodic physics, Anderson localization transition, and transport properties at the critical point etc.\cite{RevModPhys.80.1355,RevModPhys.91.021001,PhysRevB.82.174411}. Conventionally, the mobility edge is considered as a critical point between the extended and the localized states. However, some works have reported different kinds of mobility edges being found between the multifractal and the extended phases and between the multifractal and the localized phases \cite{deng2019one,10.21468/SciPostPhys.12.1.027}.

On a parallel front, the topological phases of matter offer remarkable and intriguing phenomena that aid in understanding of the crucial properties of systems \cite{RevModPhys.83.1057,RevModPhys.82.3045,RevModPhys.88.021004}. In particular and of relevance to us, the topological superconductors (TSCs) have received immense attention due to their relevance to the field of topological quantum computation \cite{Sato_2017}. The Majorana zero modes (MZMs), thought to be found in the TSCs, are considered flawless candidates for them to be used as qubits \cite{PhysRevLett.94.166802,RevModPhys.80.1083}. The unique property of the MZMs to have a great deal of importance for their non-local nature with complete localization occurring at the boundaries of the chain, and hence robust to any local perturbations \cite{Alicea_2012,RevModPhys.87.137}. A prototype theoretical model to study the TSCs and the properties of the MZMs is the Kitaev chain model \cite{Kitaev_2001}. The model describes a one-dimensional tight-binding spinless fermions in the presence of $p$-wave superconducting correlations. The MZMs are quasiparticle excitations that obey non-abelian statistics. Recently, several theoretical models have been proposed to find the signature of MZMs \cite{PhysRevB.90.014505,PhysRevB.100.064202,PhysRevB.89.115430,PhysRevB.98.165144}. From the experimental perspective, the most studied and accepted proposal is the semiconductor-superconductor hybrid systems \cite{PhysRevLett.105.177002,PhysRevB.82.214509,gul2018ballistic,mourik2012signatures,lutchyn2018majorana}. Other than that, there are some other realistic models developed as well \cite{jeon2017distinguishing,feldman2017high,ruby2017exploring}.

Among the members of the generalized Kitaev model \cite{PhysRevLett.112.206602,PhysRevLett.107.036801,PhysRevB.63.224204,PhysRevLett.110.146404,PhysRevLett.110.176403}, we are interested in a dimerized Kitaev chain, which is shown to have significant interest \cite{PhysRevB.90.014505,PhysRevB.96.205428,PhysRevB.96.121105,PhysRevB.97.085131}. A dimerized Kitaev chain is a hybrid model with a one-dimensional SSH chain and Kitaev chain,  and possess very rich physics and symmetry properties. There are signatures of a trivial phase, SSH-like, and Kitaev-like topological phases in a single system. 

Further, since disorder is an indispensable element of any quantum system, it is required to incorporate it. Thus the interplay of topology and disorder have gained a lot of attention in recent years. Up till now, it is known that a topological phase survices in presence of a weak disorder. However, there will be a transition from the topologically non-trivial phase to the topologically trivial phase in the strong disorder limit. Very recently there has been a remarkable observation where the presence of disorder can drive a trivial phase to a topological phase, known as the topological Anderson insulator\cite{PhysRevLett.102.136806}. Subsequently, several other models have reported the same behavior theoretically and experimentally \cite{PhysRevLett.103.196805,PhysRevLett.127.263004,PhysRevLett.126.146802,PhysRevLett.125.133603}. The dimerized Kitaev chain in the presence of a random potential has also demonstrated a similar behavior \cite{PhysRevB.100.205302}.

Deriving motivations from the above results, in this paper we consider a dimerized Kitaev chain in the presence of onsite QP potential. We numerically study the localization and the topological properties of this model via computing several physical quantities. We observe series of phase transitions occuring, such as extended-critical-localized phases due to the competition between the dimerization and the QP potential strength. We infer that our model hosts a critical phase consisting of two different mobility edges separating the extended and the localized phases, and a second one intervening the critical and the localized phases. Hence a broad region with critical (multifractal) states arise, resulting in an extended multifractal phase. This is a significant and a noteworthy result. In addition to this, we study the topological properties of the zero-energy edge modes. We find that the onsite QP potential will drive the system from a topologically trivial to a non-trivial (Topological Anderson) phase beyond a certain critical dimerization strength. Beyond this critical strength, it will exhibit another transition from a topologically non-trivial phase to the Anderson localized phase in presence of a strong QP potential.

The rest of the paper is organized as follows. First, we describe the model in section II. Further, the results are reported and analyzed in section III. Here, we discuss the localization and topological properties in subsections A and B, respectively. Finally, we conclude our observations in section IV.

\section{Model}\label{model}
Here we consider a one-dimensional spinless fermionic chain comprising of two distinct atoms (sublattices) in a unit cell. The hopping and the $p$-wave superconducting pairing strengths are assumed to alternate between strong (within the unit cell) and weak bonds (between the unit cells). Moreover, the onsite chemical potentials ($\mu$) at the two sublattices within a unit cell are modulated quasiperiodically. The Hamiltonian of such a system is represented by,

\begin{align}
H=-&t\sum_{m=1}^{N}\bigg((1+\delta) \hat{c}^{\dagger}_{m,B}\hat{c}_{m, A}+{\rm {H.c.}}\bigg)\nonumber \\
-&t \sum_{m=1}^{N-1}\bigg( (1-\delta) \hat{c}^{\dagger}_{m+1, A}\hat{c}_{m, B}+ {\rm {H.c.}}\bigg) \nonumber\\
+&\Delta\sum_{m=1}^{N}\bigg((1+\delta) \hat{c}^{\dagger}_{m,B}\hat{c}^{\dagger}_{m, A}+{\rm {H.c.}}\bigg)\nonumber \\
+&\Delta \sum_{m=1}^{N-1}\bigg( (1-\delta) \hat{c}^{\dagger}_{m+1, A}\hat{c}^{\dagger}_{m, B}+ {\rm {H.c.}}\bigg) \nonumber \\
-&\sum_{m=1}^{N}[\mu _{A}c^{\dagger}_{m,A} c_{m,A}+\mu _{B} c^{\dagger}_{m,B} c_{m,B})
\label{eqn:ham}
\end{align}
where the quasiperiodically modulated onsite chemical potentials at the two sublattices are denoted by,
$$\mu_{A}=\lambda_{A}\cos[2\pi\beta(2m-1)+\phi]$$
$$\mu_{B}=\lambda_{B}\cos[2\pi\beta(2m)+\phi].$$
Here, the length of the chain is represented by $L=2N$ with the unit cell index $m~(=1,~2,~,~,N)$. In each unit cell, there are two sublattice sites, namely, $A$ and $B$ the corresponding number operators being $\hat{n}_{m,A}$ and $\hat{n}_{m,B}$, respectively. The creation (annihilation) operators to create an electron at the sublattice sites $(m,A)$ and $(m,B)$ are given by $\hat{c}^{\dagger}_{m,A}~(\hat{c}_{m, A})$ and $\hat{c}^{\dagger}_{m, B}~(\hat{c}_{m, B}$), respectively. The intracell (strong) and intercell (weak) hopping strengths are defined by $t(1+\delta)$ and $t(1-\delta)$ with $t$ and $\delta$ being the nearest-neighbour hopping strength and the dimensionless dimerization strength, respectively. To keep the hopping term positive, we impose a constraint on the dimerization strength, namely, $|\delta|<1$.
Similarly, the intracell (strong) and intercell (weak) $p$-wave superconducting paring strength of the system are defined by $\Delta(1+\delta)$ and $\Delta(1-\delta)$.  
The onsite quasiperiodic potential strengths at the two sublattices are denoted by $\lambda_{A}$ and $\lambda_{B}$, respectively. The periodicity of the potential is given by $1/\beta$. In this work, $\beta$ is taken as the golden ratio, that is, $\beta=\frac{(\sqrt{5}-1)}{2}$. The phase term of the potential is represented by $\phi$ which is taken as zero. We keep the $p$-wave pairing strength to be real and positive, that is, $\Delta=0.5$. Additionally, we choose the onsite quasiperiodic potential strengths $\lambda_{A}$ and $\lambda_{B}$ to be equal and opposite in magnitude, that is, $\lambda_{A}=-\lambda_{B}=\lambda$. We have taken $t$ as the unit of energy throughout.

In the presence of a $p$-wave superconducting pairing, the Hamiltonian has terms quadratic in the fermionic creation (and annihilation) operators. Thus, the Hamiltonian can be solved by using the Bogoliubov-de Gennes (BdG) transformation.
The particle-hole symmetry is inherently present in the BdG Hamiltonian. Thus corresponding to each particle-like solution ($u_{n},v_{n}$) with eigenenergies $+E$, there will be a hole-like solution with $-E$. Only the zero-energy states ($E=0$) are self-conjugate.

In this work, we shall study the effect of an onsite QP potential on the localization properties via analysing the eigenenergies and the eigenstates of the Hamiltonian, and explore the topological properties using the real-space winding number and the number of the of MZMs present in the system. The results will elucidate the critical properties of the model and shed light on the topological properties as well. 

\section{Results}

\begin{figure}[!t]
\centerline{\hfill
\includegraphics[width=0.5\textwidth]{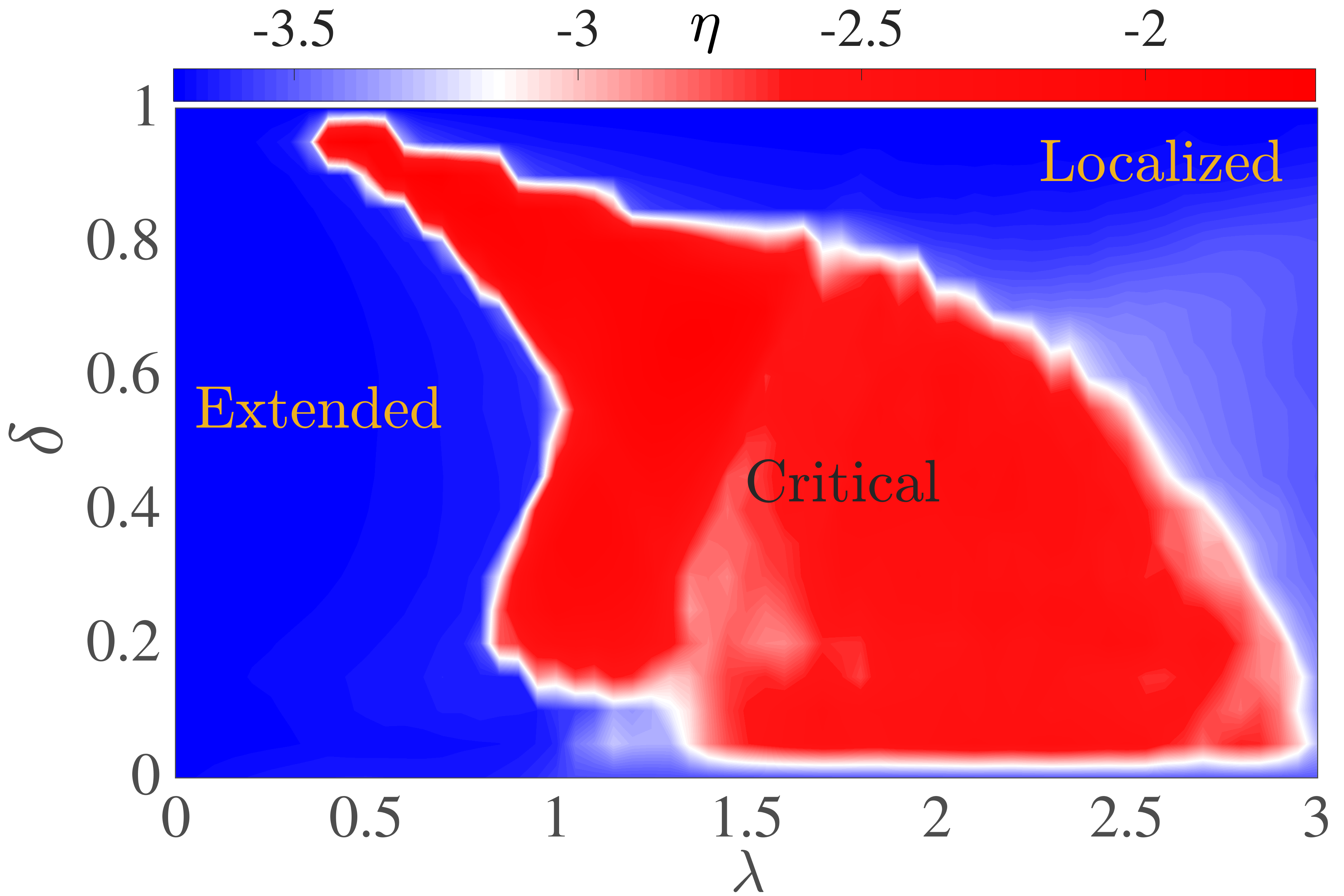}
\hfill}
\caption{The phase diagram is shown using the variable $\eta$ as a function of $\delta$ and $\lambda$. The length of the chain we consider for the calculation is $L=5168$.}
\label{fig:eta_phasediag}
\end{figure}

\subsection{Localization Study}
In this section, we explore the localization properties of the eigenstates of the system under periodic boundary condition. For this section, we use two diagnostic tools, such as inverse participation ratio (IPR) and the normalized participation ratio (NPR) to distinguish between the extended, critical, and localized properties. The IPR and the NPR corresponding to the $n^{th}$ eigenstate of the BdG Hamiltonian and defined as \cite{PhysRevLett.126.106803},
\begin{equation}
{\rm{IPR}}^{(n)}=\sum_{m=1,\alpha=(A,B)}^{N} (|u^{(n)}_{m,\alpha}|^{2}+v^{(n)}_{m,\alpha}|^{2})^{2}
\end{equation}
and
\begin{equation}
{\rm{NPR}}^{(n)}=\bigg[\it{L}~\sum_{m=1, \alpha=(A,B)}^{\it{N}} (|u^{(n)}_{m,\alpha}|^{2}+v^{(n)}_{m,\alpha}|^{2})^{2} \bigg] ^{-1}
\end{equation} 
where $u^{(n)}_{m,\alpha}$ and $v^{(n)}_{m,\alpha}$ are the solutions of the BdG equations.
It is known that the IPR value of an extended state goes to zero, while for the localized state, it is always stays finite and acquires a value '$1$' in the thermodynamic limit. On the other hand, the NPR value denotes finite values corresponding to an extended state, while, for the localized state, it tends to zero in the thermodynamic limit. 
Moreover, we are interested in the global properties of the model. Since our system respects the particle-hole symmetry, we only consider the upper half of the energy spectrum in our calculations. Hence, the average of the IPR and the NPR over the upper half of the total number of eigenstates of the energy spectrum are given by \cite{PhysRevLett.126.106803},
\begin{eqnarray}
\langle{\rm{IPR}}\rangle &=\frac{1}{L}\sum_{n=1}^{L} {\rm{IPR}}^{(n)}\\  \nonumber
\langle{\rm{NPR}}\rangle &=\frac{1}{L}\sum_{n=1}^{L} {\rm{NPR}}^{(n)}.
\label{av_IPR_NPR}
\end{eqnarray}

\begin{figure}[!t]
\centerline{\hfill
\includegraphics[width=0.25\textwidth]{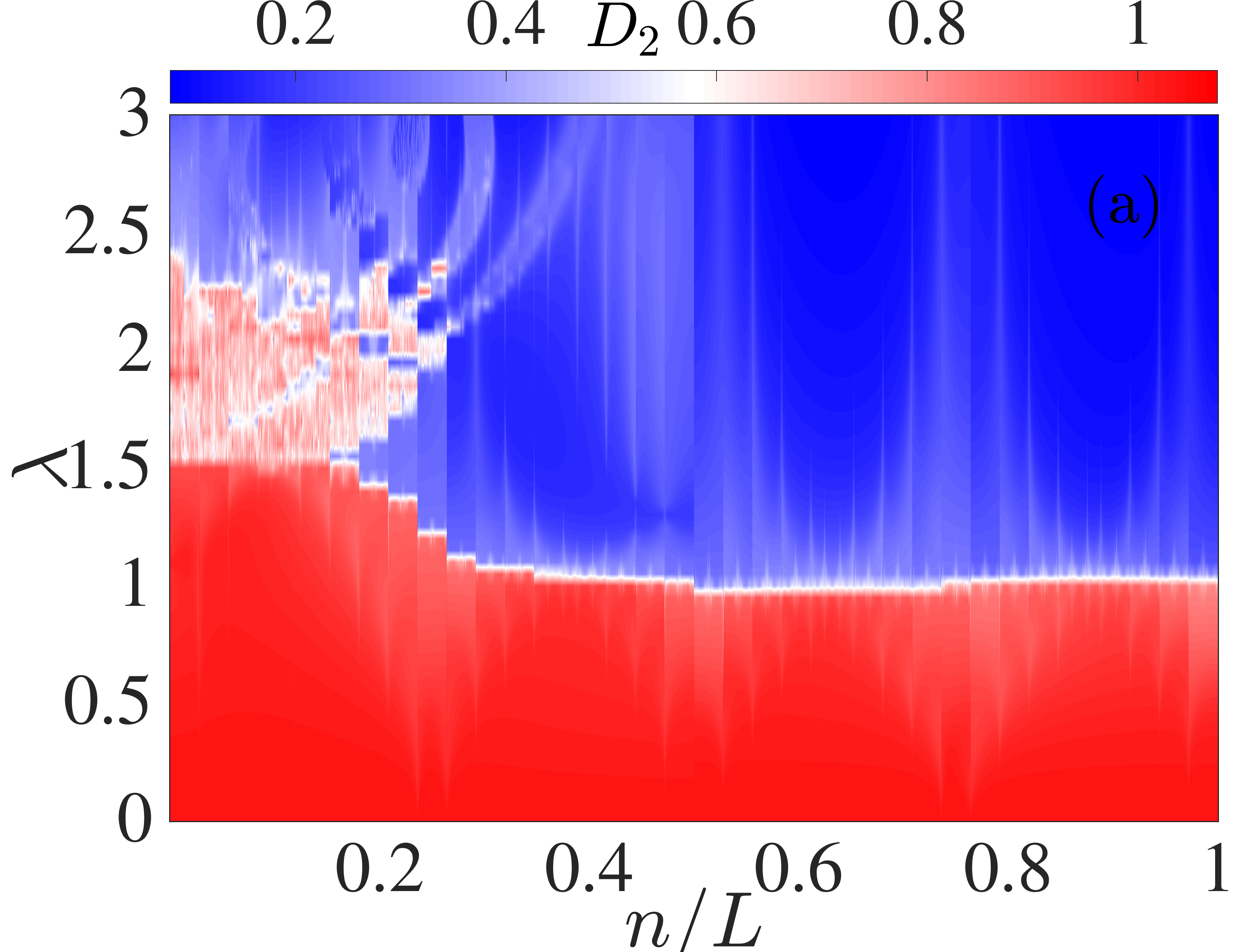}
\hfill
\hfill
\includegraphics[width=0.25\textwidth]{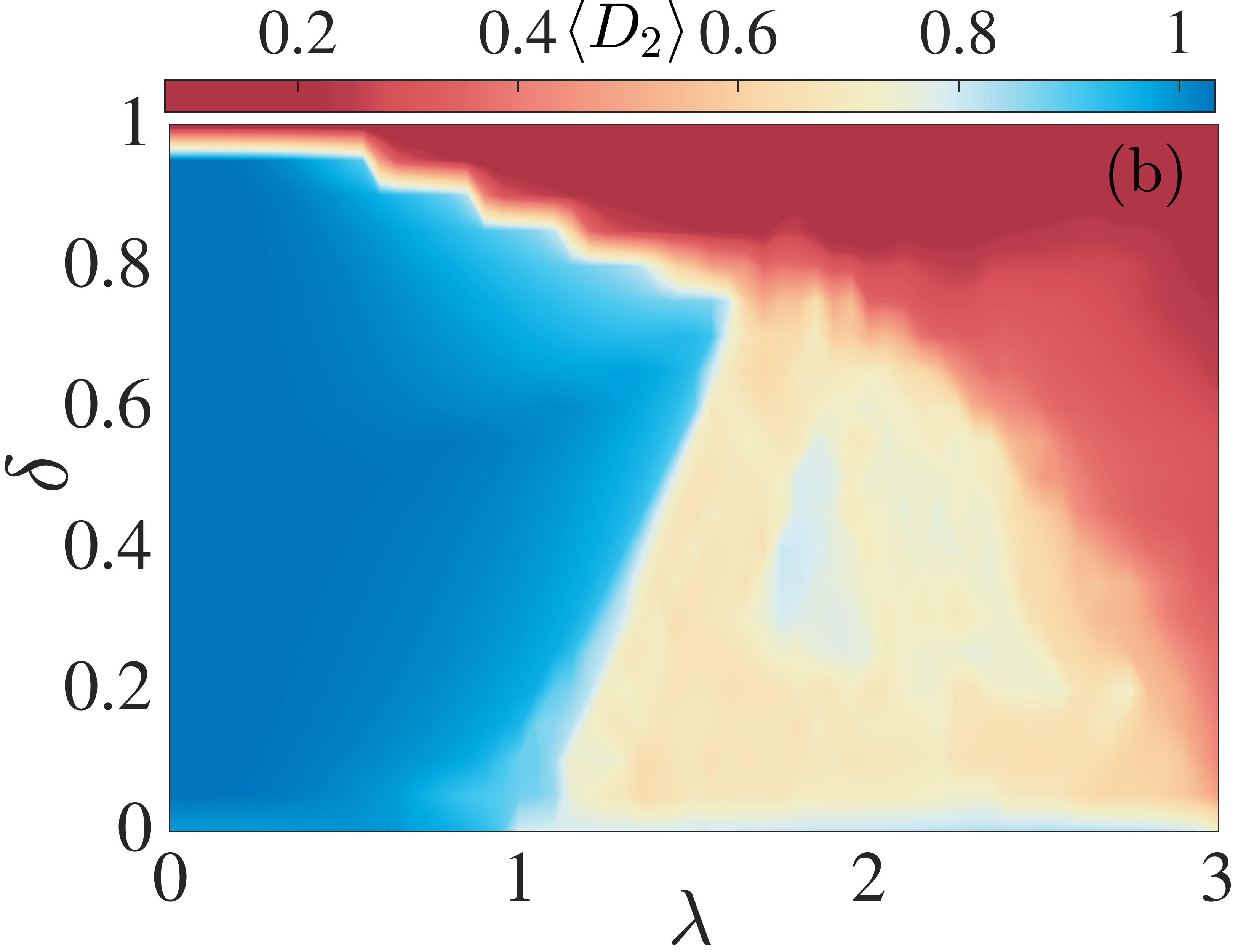}
\hfill}
\caption{(a) The fractal dimensions ($D_{2}$) are shown for the upper half of the energy states as a function of $\lambda$. The system length is taken as $L=8362$. (b) A phase diagram using the fractal dimensions ($D_{2}$) is plotted as a function of $\delta$ and $\lambda$. In this calculation, we consider only a band of lower energy states from the upper half of the energy spectrum. The length of the chain we consider for the calculation is $L=5168$. }
\label{fig:D2_phasediag}
\end{figure}

\begin{figure*}[!t]
\centerline{\hfill
\includegraphics[width=0.33\textwidth]{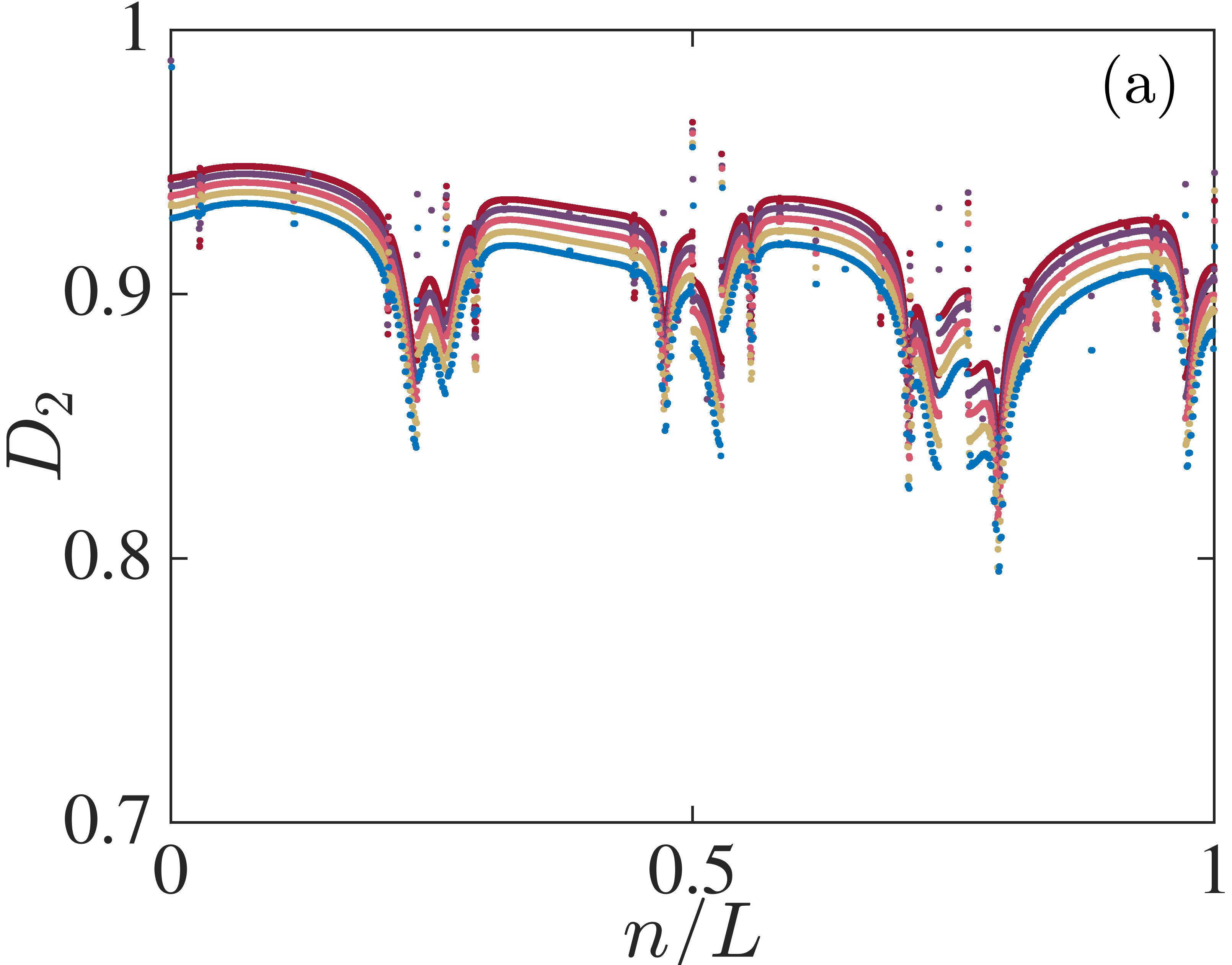}
\hfill
\hfill
\includegraphics[width=0.35\textwidth]{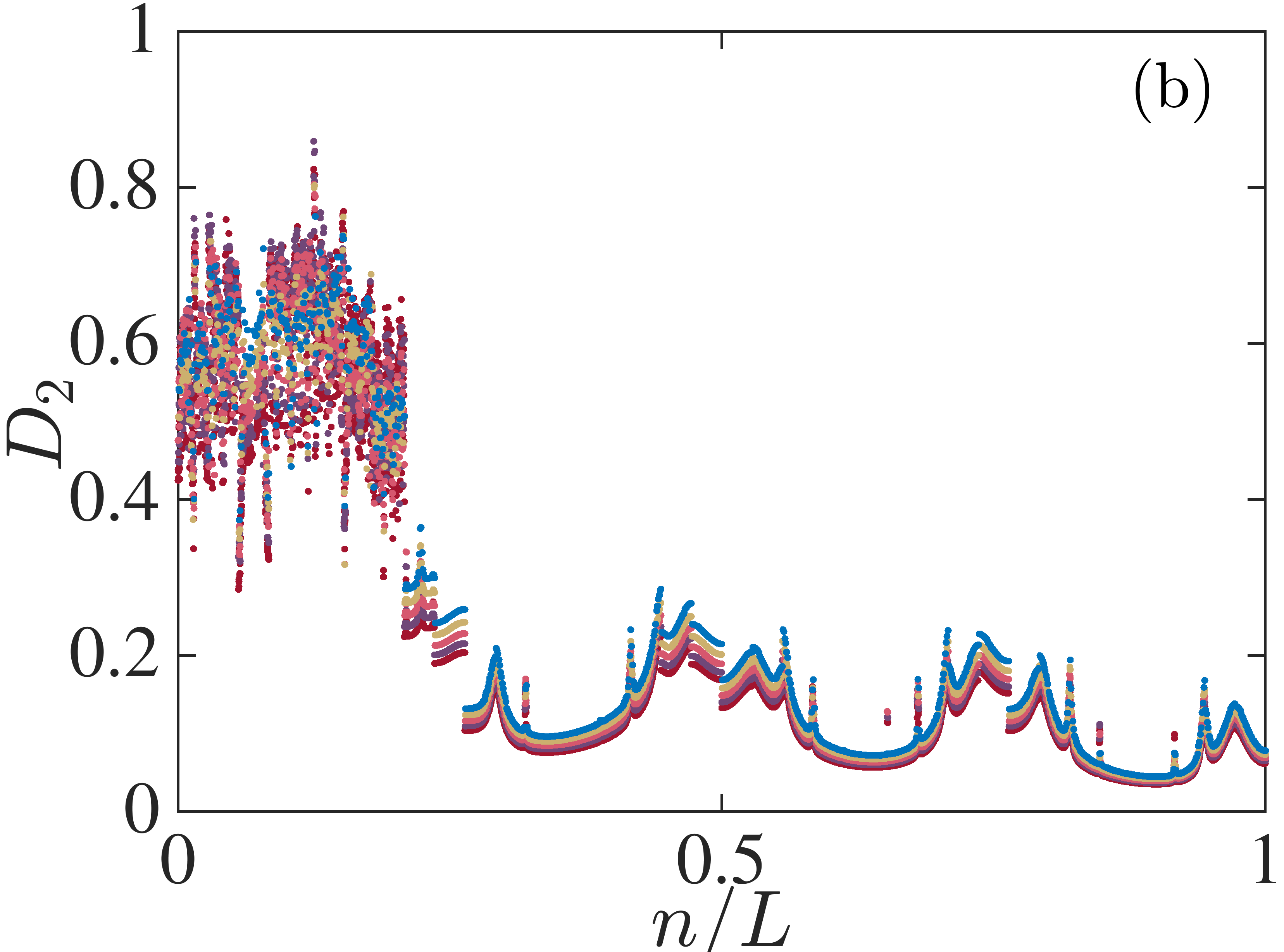}
\hfill
\hfill
\includegraphics[width=0.34\textwidth]{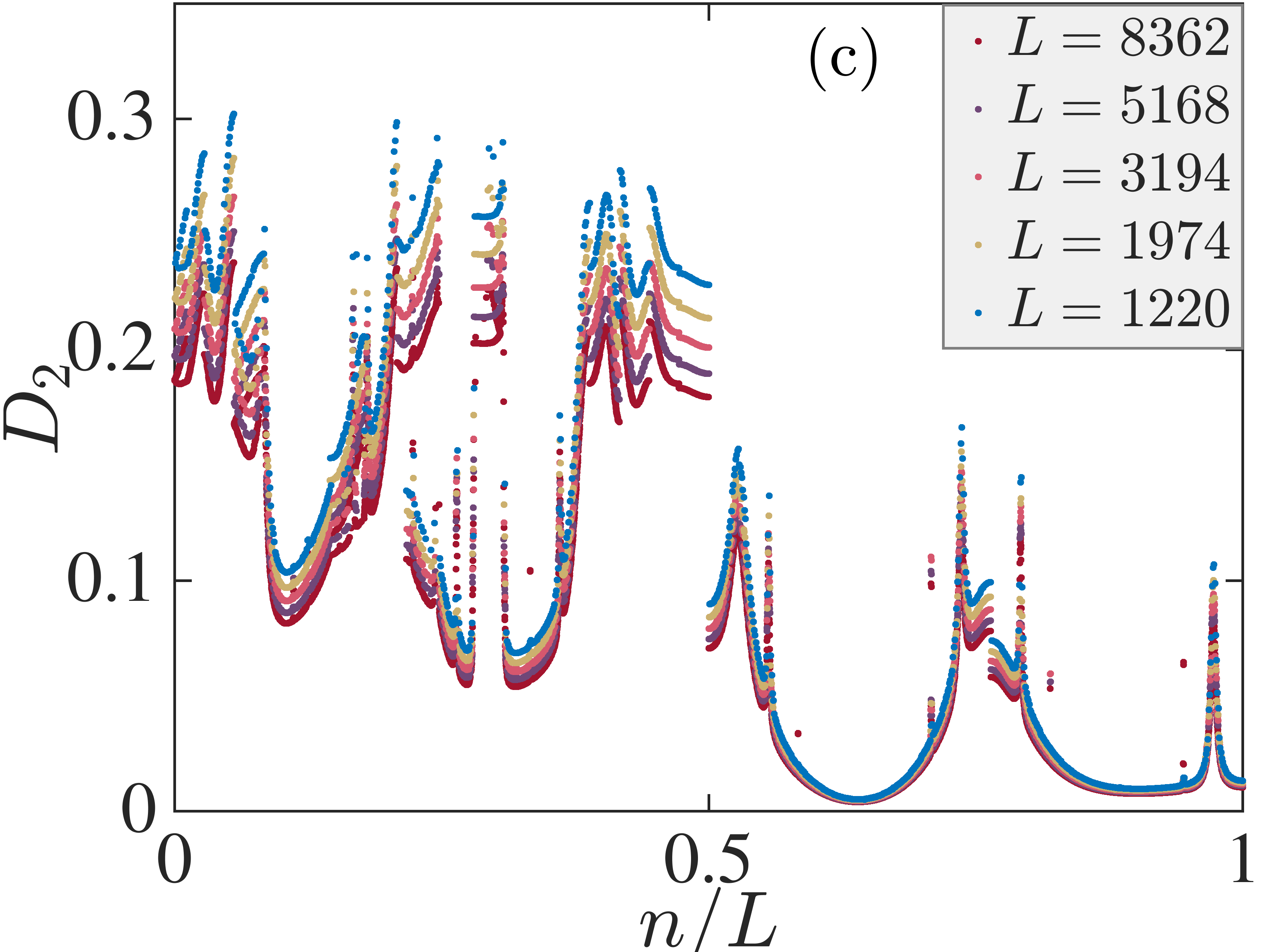}
\hfill}
\caption{The fractal dimension ($D_{2}$) as a function of eigenstate index ratio ($n/L$) ( upper half of the energy spectrum ) are shown in (a) $\lambda=0.50$, $\delta=0.60$, (b) $\lambda=1.70$, $\delta=0.60$, and (c) $\lambda=3.00$, $\delta=0.60$ for various system sizes, mentioned in the figure.}
\label{fig:D2_eigenstateindex}
\end{figure*}

\begin{figure*}[!t]
\centerline{\hfill
\includegraphics[width=0.34\textwidth]{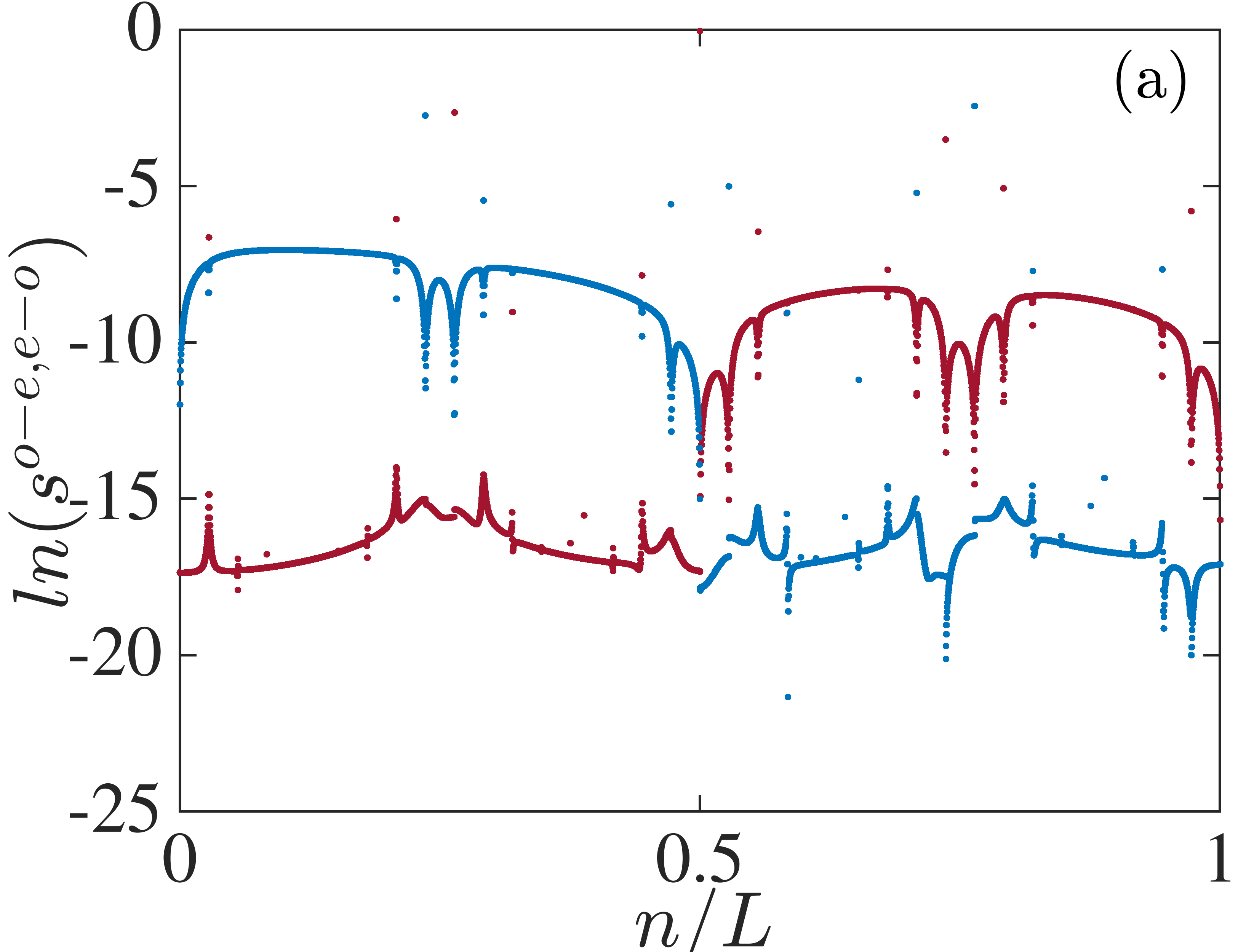}
\hfill
\hfill
\includegraphics[width=0.335\textwidth]{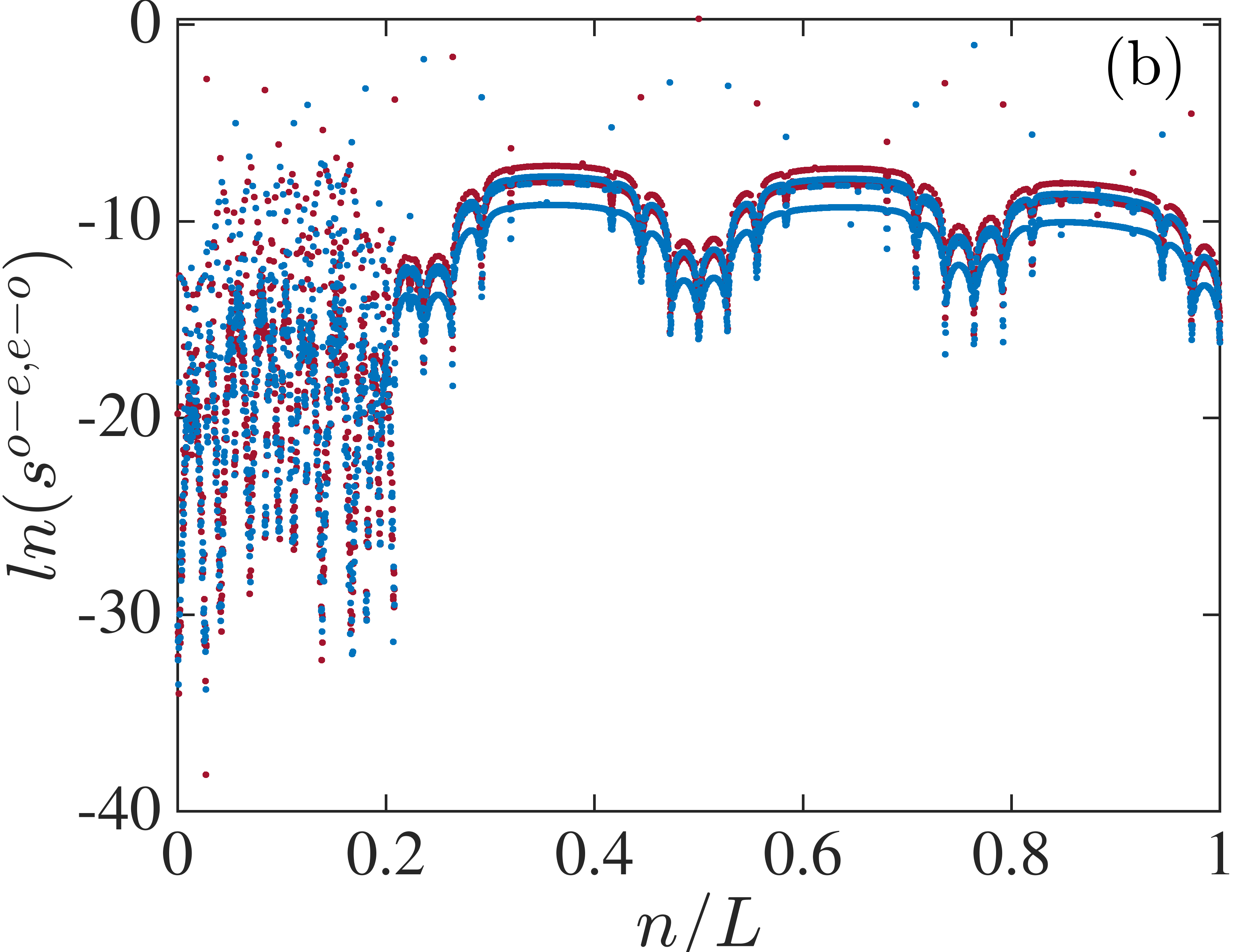}
\hfill
\hfill
\includegraphics[width=0.34\textwidth]{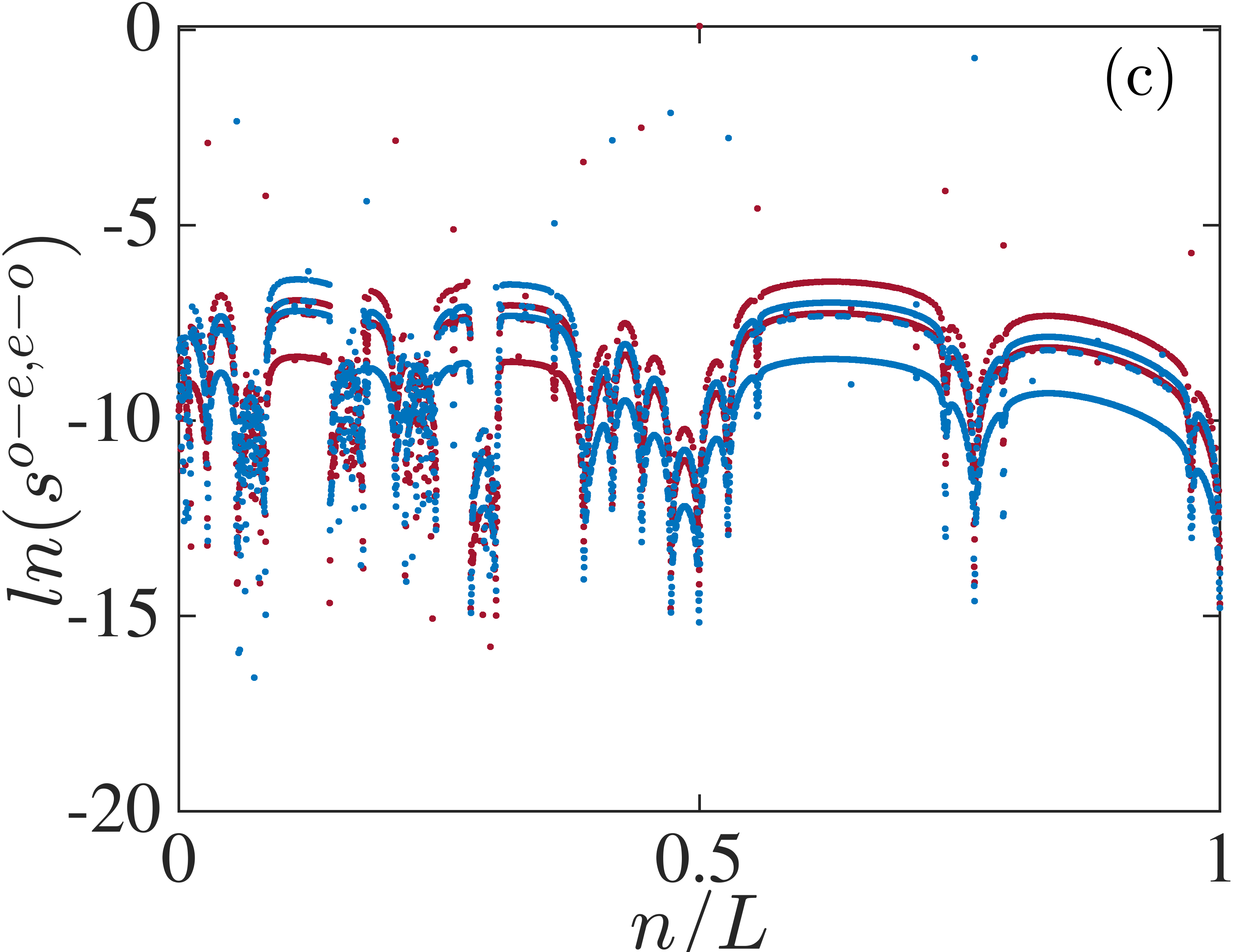}
\hfill}
\caption{The level spacing as a function of eigenstate index ratio ($n/L$) ( upper half of the energy spectrum ) are shown in (a) $\lambda=0.50$, $\delta=0.60$, (b) $\lambda=1.70$, $\delta=0.60$, and (c) $\lambda=3.00$, $\delta=0.60$. The system length is taken as $L=8362$.}
\label{fig:level_spacing}
\end{figure*}

To begin with, we present a phase diagram with the help of $\langle{\rm{IPR}}\rangle$ and $\langle{\rm{NPR}}\rangle$ in the parameter space defined by the dimerization strength ($\delta$) and the QP potential strength ($\lambda$) in Fig~\ref{fig:eta_phasediag}. In order to obtain a detailed illustration of the phase diagram, we need to segregate different phases, such as the extended, critical (intermediate), and the localized phases. The coexistence of different types of states (phases) gives rise to the critical phase. To this end, we calculate a quantity, $\eta$ which is given by \cite{PhysRevLett.126.106803},
\begin{equation}
\eta={\rm{log}}_{10}{({\langle{\rm{IPR}}\rangle}  \times  {\langle{\rm{NPR}}\rangle})}.
\label{eta}
\end{equation}
The value of $\eta$ distinguishes the critical phase from the extended and the localized ones. However, it is incapable of distinguishing between the localized phase from the delocalized phase. Thus, it only helps us to identify the critical phase in the phase diagram. Hence, we need another quantity to discern the localized phase from the delocalized phase.

The fractal dimension is an excellent quantity to identify different phases accurately. The fractal dimension, $D_{2}$ is defined as \cite{deng2019one,yao2019critical},
\begin{equation}
D_{2}=-\lim_{L\rightarrow \infty} \frac{log{\rm{(IPR)}}}{log(L)}.
\label{D2}
\end{equation} 
While it has a value $1$($0$) for an extended(localized) state, a critical/multifractal state in the thermodynamic limit will have a value in between $0$ and $1$. Following this, the average value of the fractal dimension calculated over a narrow band comprising of a few states and upper half of all the states are denoted by $\langle D_{2} \rangle$ and $\overline{D_{2}}$, respectively.   
Again, the average value of the fractal dimension will not capture the overall nature of the system. Thus, we need to consider both the quantities, namely, $\eta$ and the average value of the fractal dimension $\langle D_{2}\rangle $ together to acquire a good knowledge on the emergent phases of the system.

In Fig~\ref{fig:eta_phasediag}, we show the phase diagram using $\eta$ in the parameter space spanned by $\delta$ (dimerization strength) and $\lambda$ (QP potential). It denotes the global nature of the system, that is including the upper half of the states, with the 'Blue' color corresponding to the extended and the localized phases and the 'red' color refers to the critical phase of the system. Therefore, the system hosts a critical phase over a large parameter regime denoted by $\delta$ and $\lambda$. Among the two extreme cases, that is, when no dimerization is present ($\delta=0$), it is observed that, all the single-particle states are extended in nature up to a value $\lambda \simeq 1$. Upon increasing the potential strength, the eigenstates become critical (multifractal). On the contrary, in the strong dimerization limit ($\delta=1$), all the states are localized irrespective of the values of $\lambda$. Further, at an intermediate point of the dimerization strength, say, $\delta \simeq 0.6$, it is observed that an extended phase persists up to $\lambda\sim 1$, beyond which, a critical phase appears which persists up to a value given by $\lambda_{c} \simeq 2.4$. Finally, the localization transition occurs at values larger than the critical $\lambda_{c}$, leading to a completely localized phase.

In Fig~\ref{fig:D2_phasediag} (a), we show an intuitive picture of the eigenspectra and their sensitivity to the variation of $\lambda$ for $\delta=0.6$. Hence, we plot $D_{2}$ corresponding to the upper half of the energy spectrum as a function of the QP potential strength, $\lambda$. Different eigenstates experience localization transitions at different values of the potential  indicating the presence of an energy-dependent phase transition. Thus a mobility edge should be observed in the presence of the dimerization and the staggered potential. In general, it is observed that the lower energy states of the spectrum (near the zero energy) are necessary to demonstrate a localization transition at large values of $\lambda$. In comparison, the higher energy states undergo a transition at weaker potential strengths. For small values of $\lambda$, all the single particle eigenstates are extended in nature, thereby giving rise to a completely delocalized phase. Beyond this, the onset of localization occurs at $\lambda \simeq 1$ corresponding to of the higher energy states, leading to a critical phase comprising of a mixture of the extended and the localized states. Hence, a mobility edge appears between the extended and the localized states. Interestingly, while with the increase in $\lambda$, the critical phase persists, the extended nature corresponding to the lower energy states is replaced by the critical (multifractal) states within a range of $\lambda$ given by $1.5 <\lambda<2.5$. Thus we observe another mobility edge, which arises between the critical (multifractal) and the localized states. Finally, at a higher value of $\lambda$, all states become localized.

Therefore, we infer that the lower energy eigenstates experience a series of transitions, namely, from the extended to critical (multifractal) and hence to the localized one. This indicates the presence of a critical region sandwiched between the extended and the localized phases corresponding to the states at lower energies. Although, corresponding to the higher energy states, there is a sharp transition from an extended to a localized phase. Thus, our results offer two different mobility edges, one between the extended and the localized and another between the critical (multifractal) and the localized phases. Hence obtaining two different mobility edges in the dimerized Kitaev chain model in presence of a QP potential comprises of an important highlight of our work.

In Fig~\ref{fig:D2_phasediag} (b), we show the average value of the fractal dimension ($\langle D_{2} \rangle$) in the parameter space spanned by of $\delta$ and $\lambda$ corresponding to a band of lower energy states appearing in Fig~\ref{fig:D2_phasediag} (a). It is depicted that in the case of weaker potential strengths (small $\lambda$), all the eigenstates are extended in nature irrespective of the value of $\delta$ ($0<\delta<1$). Beyond the critical point, $\lambda \simeq 1$, a critical phase appears with critical (multifractal) nature of the eigenstates within $0.2 < \delta <0.8$. These multifractal states are affected by larger values of $\delta$, resulting in shrinking of the critical phase. Finally, a complete localization occurs at stronger potential strengths. Therefore, we observe three distinct phases, such as, the extended, critical (multifractal), and the localized as a function of the QP potential strength $\lambda$. 

In order to have a complete understanding of these different phases, we study $D_{2}$ by considering different system sizes, such that, $L=8362,~5168,~3194,~1974,$ and $1220$, which are shown via different colors in Fig~\ref{fig:D2_eigenstateindex}. In Fig~\ref{fig:D2_eigenstateindex}, we plot $D_{2}$ as a function of the eigenstate index ratio ($n/L$) corresponding to three representative points (parameter values) from each of the phases. For the dimerization strength, $\delta=0.6$, we choose $\lambda=0.5$ for the extended phase, $\lambda=1.7$ for the critical (multifractal) phase, and $\lambda=3.0$ for the localized phase. In Fig~\ref{fig:D2_eigenstateindex} (a), it is observed that, the values of $D_{2}$ corresponding to all the states move towards the value $D_{2}=1$ as $L$ increases, implying a the presence of a completely extended phase in the thermodynamic limit. Most interestingly, in Fig~\ref{fig:D2_eigenstateindex} (b), the values of $D_{2}$ fluctuate around a value $D_2\simeq 0.6$ for different $L$, thereby demonstrating a fractal nature. However, the higher energy states approach towards a value $D_{2}=0$ with increasing $L$, indicating a localized behavior. Although the mobility edge refers to the critical point of the extended and the localized states, here we observe the mobility edge to occur between the multifractal and the localized states. Finally, in Fig~\ref{fig:D2_eigenstateindex} (c), we find that all the states corresponding to both the lower and the higher energies approach zero with increasing $L$, thereby exhibiting a completely localized phase.

\begin{figure}[!t]
\centerline{\hfill
\includegraphics[width=0.5\textwidth]{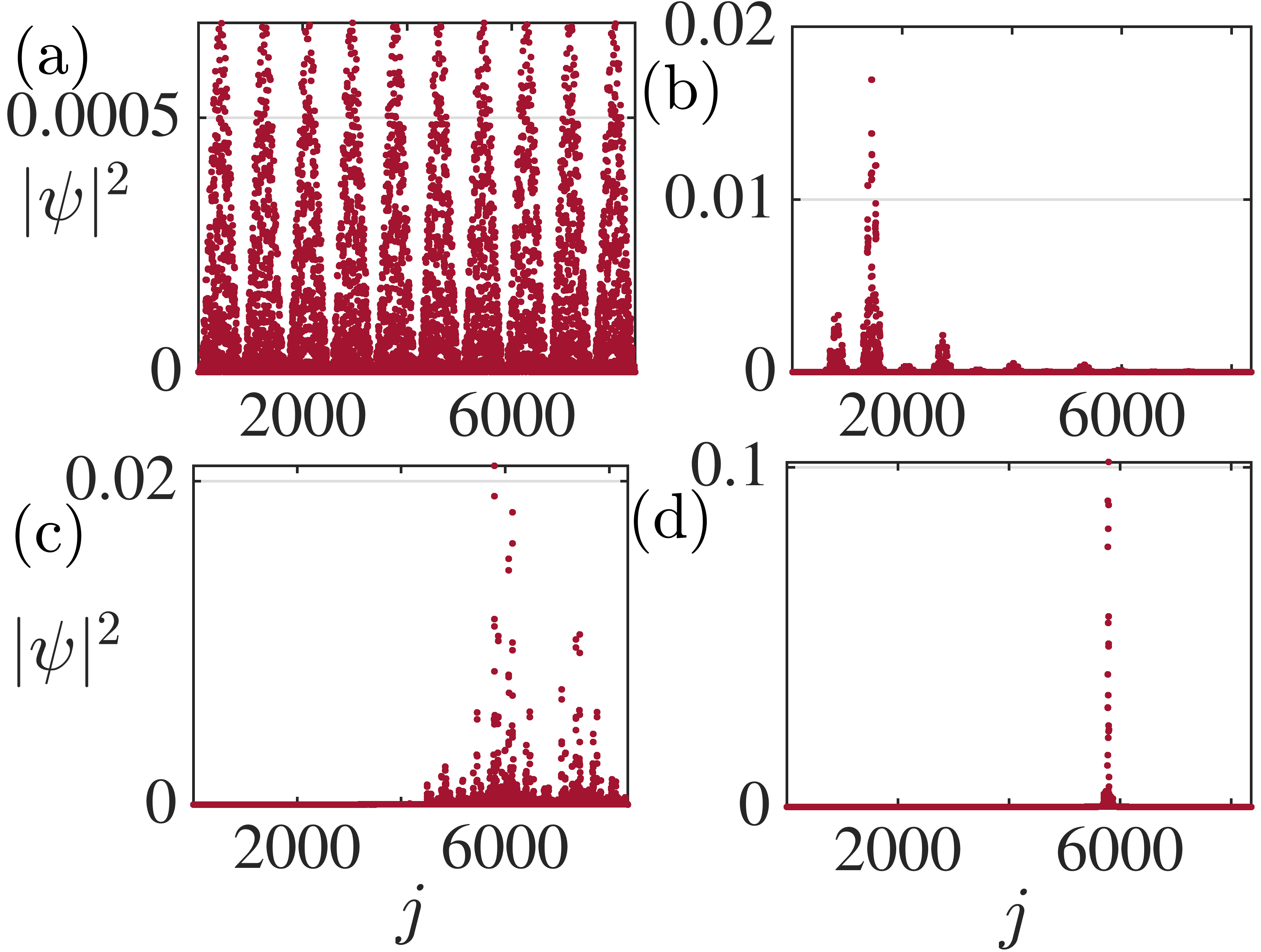}
\hfill}
\caption{The probability distribution of eigenstates as a function of site indices ($j$) are shown for to (a) $\lambda=1.5$, (b) $\lambda=1.55$, (c) $\lambda=2.4$, and (d) $\lambda=2.45$ corresponding to $\delta=0.6$. The system size we consider here is $L=8362$. }
\label{fig:bulk_prob}
\end{figure}

\begin{figure*}[!t]
\centerline{\hfill
\includegraphics[width=0.33\textwidth]{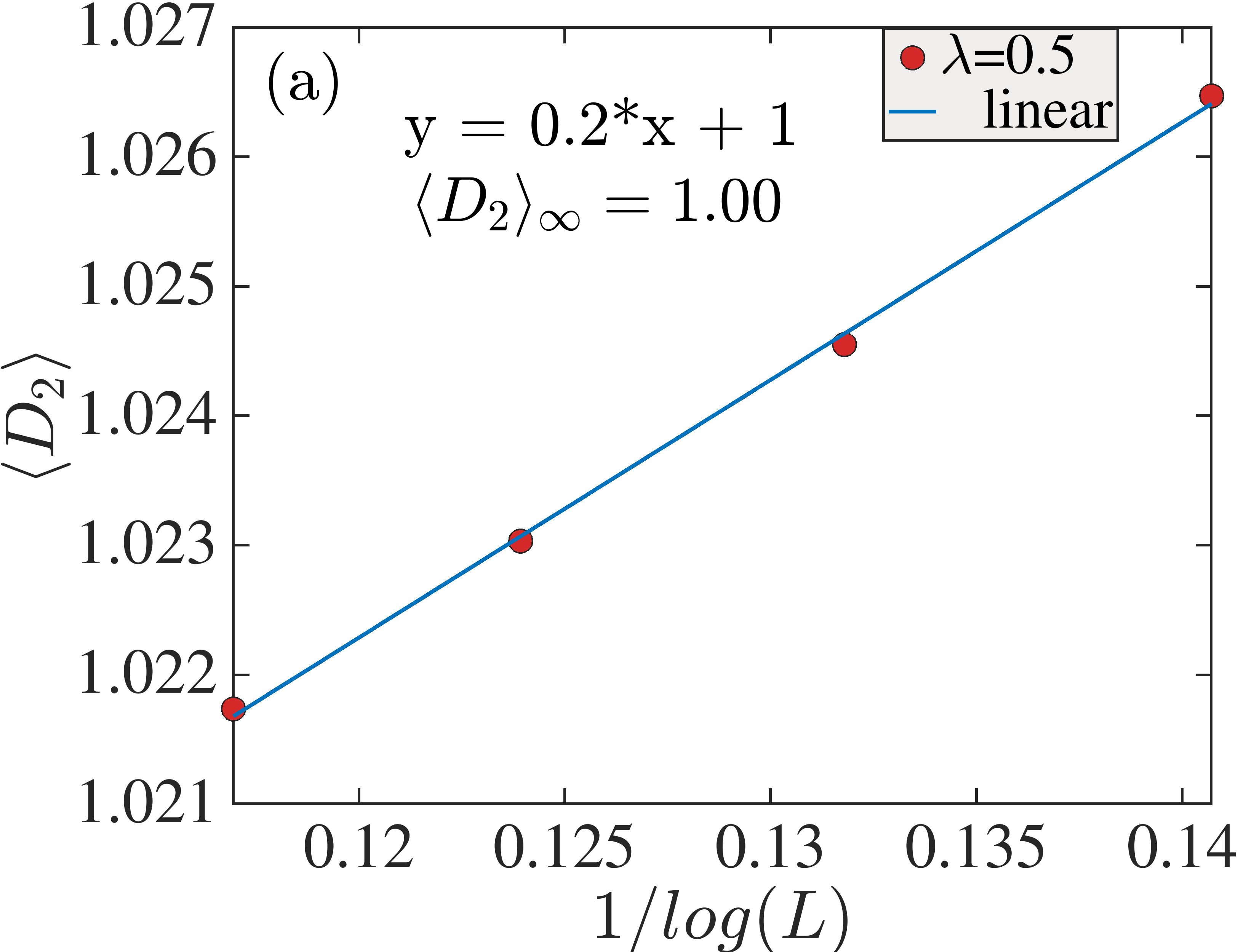}
\hfill
\hfill
\includegraphics[width=0.33\textwidth]{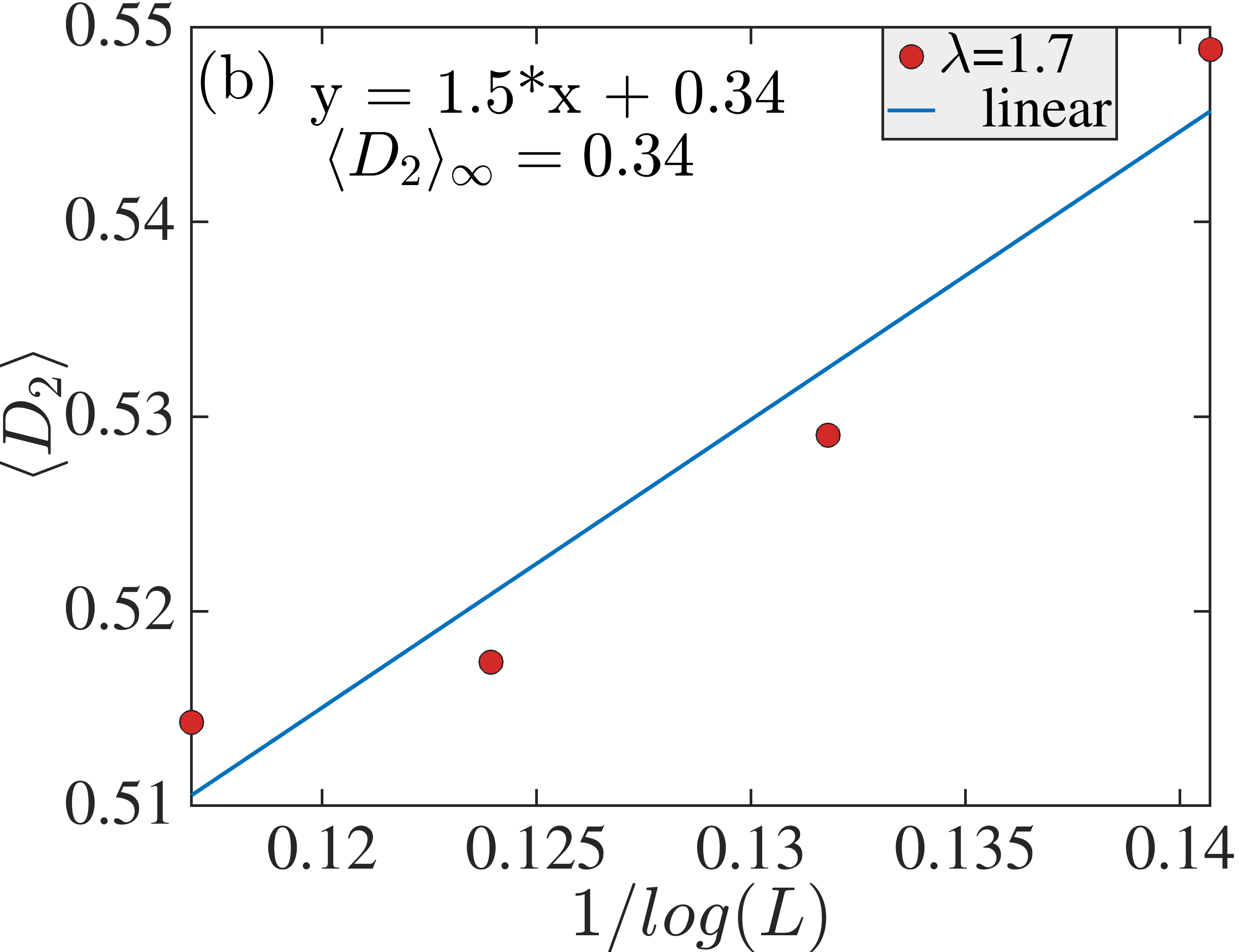}
\hfill
\hfill
\includegraphics[width=0.33\textwidth]{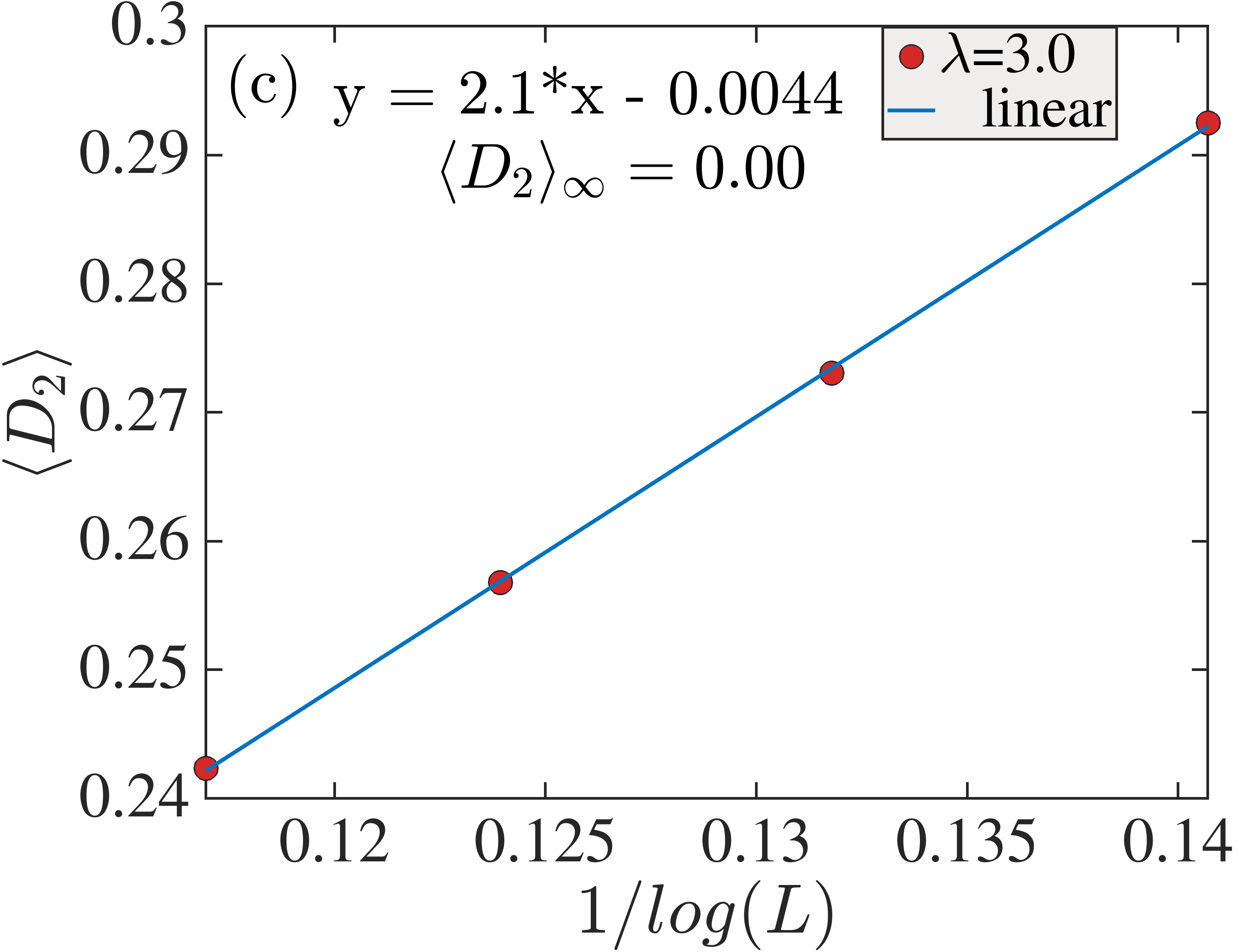}
\hfill}
\caption{The average value of fractal dimension ($\langle D_{2} \rangle$) as a function of system sizes are shown in (a) $\lambda=0.50$, $\delta=0.60$, (b) $\lambda=1.70$, $\delta=0.60$, and (c) $\lambda=3.00$, $\delta=0.60$. The system sizes are taken as $L=5168,~3194,~1974,$ and $1220$. }
\label{fig:D2_scaling}
\end{figure*}

\begin{figure}[!b]
\centerline{\hfill
\includegraphics[width=0.25\textwidth]{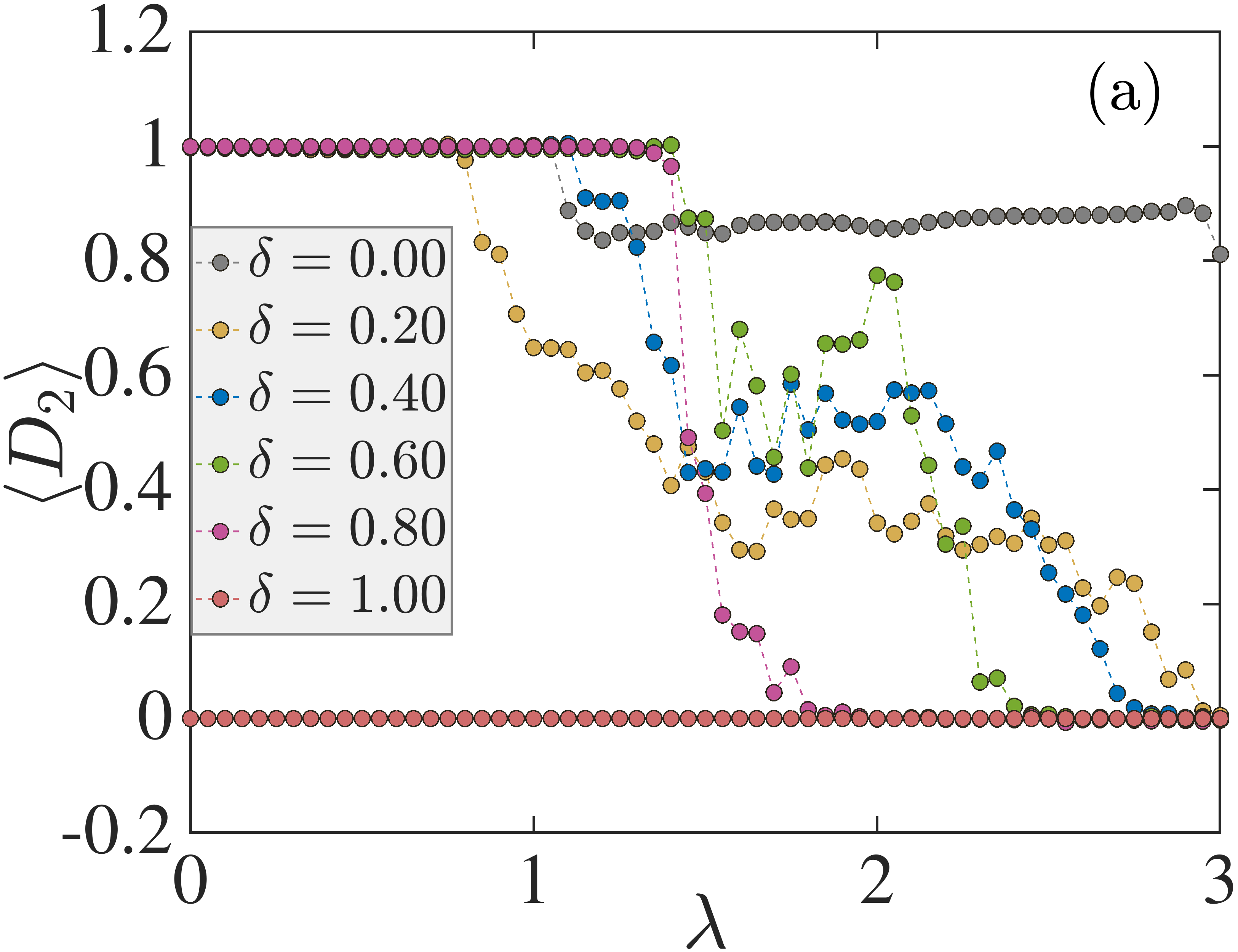}
\hfill
\hfill
\includegraphics[width=0.25\textwidth]{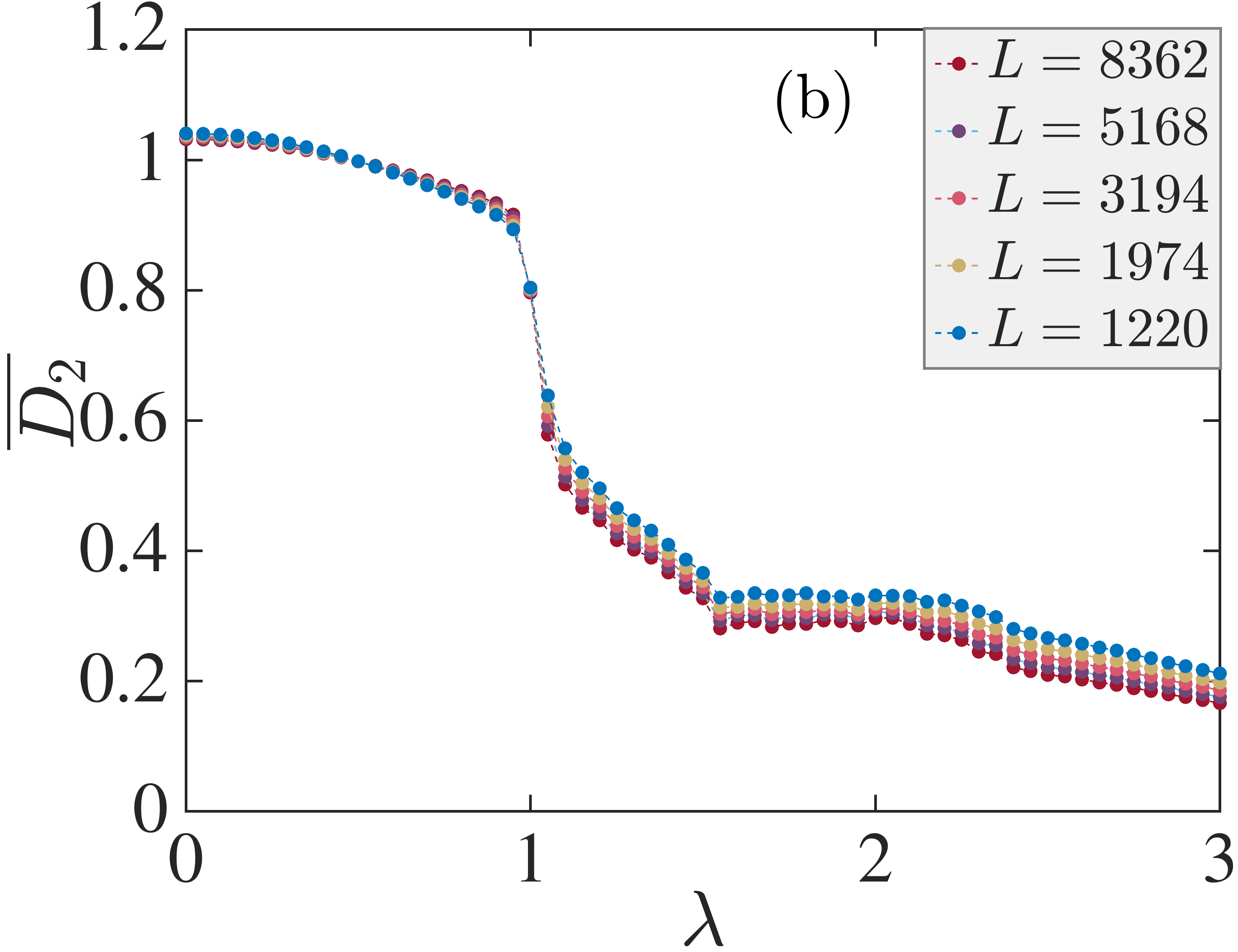}
\hfill}
\caption{(a) The average  value of $D_{2}$ ($\langle D_{2} \rangle$) over a narrow band of states and (b) The average  value of $D_{2}$ ($\overline{ D_{2}}$) over entire upper half of states as a function of $\lambda$ are shown.   }
\label{fig:D2_avD2_lambda}
\end{figure}

Generally, the fractal dimension carries the information about the eigenstates of the system. Hence, we complement our results by analyzing another quantity, namely, the energy level spacing, which uses the eigenenergies. In this calculation, we use the energies corresponding to the upper-half of the spectrum. The energy eigenvalues are arranged in ascending order, that is, $E_{1} < E_{2}<..<E_{L}$. Now, corresponding to a given energy $E_{n}$ with $n=1,~2,~.~.~.L$, the even-odd and the odd-even spacings can be calculated via \cite{deng2019one},
\begin{equation}
s^{e-o}_{n}=E_{2n}-E_{2n-1}
\end{equation}
\begin{equation}
s^{o-e}_{n}=E_{2n+1}-E_{2n}.
\end{equation}
Due to the presence of doubly degenerate eigenvalues in the extended phase, the value of $s^{e-o}_{n}$ will be non-zero, while $s^{o-e}_{n}$ will not be zero. Thus a gap will occur in the spectrum. On the other hand, there will be no gap in the localized phase. However, a distribution of fluctuations for $s^{e-o}_{n}$ and $s^{o-e}_{n}$ will be present corresponding to the critical phase. 

In Fig~\ref{fig:level_spacing} we plot the level-spacing corresponding to the same parameter choices as that for the calculation of $D_{2}$ in Fig~\ref{fig:D2_eigenstateindex}, which are $\lambda=0.5,~1.7$, and $3.0$ for $\delta=0.6$. In this study, we expect to witness a gap in the extended phase corresponding to $\lambda=0.5$ shown in Fig~\ref{fig:level_spacing} (a). Later, for $\lambda=1.7$ in Fig~\ref{fig:level_spacing} (b), distinctly noticeable fluctuations occur in the lower energy spectrum. While the higher energy states are localized in nature. Further, at $\lambda=3$ in Fig~\ref{fig:level_spacing} (c), no gap between $s^{e-o}_{n}$ and $s^{o-e}_{n}$ is observed thereby indicating a localized behavior.

\begin{figure}[!b]
\centerline{\hfill
\hfill
\includegraphics[width=0.5\textwidth]{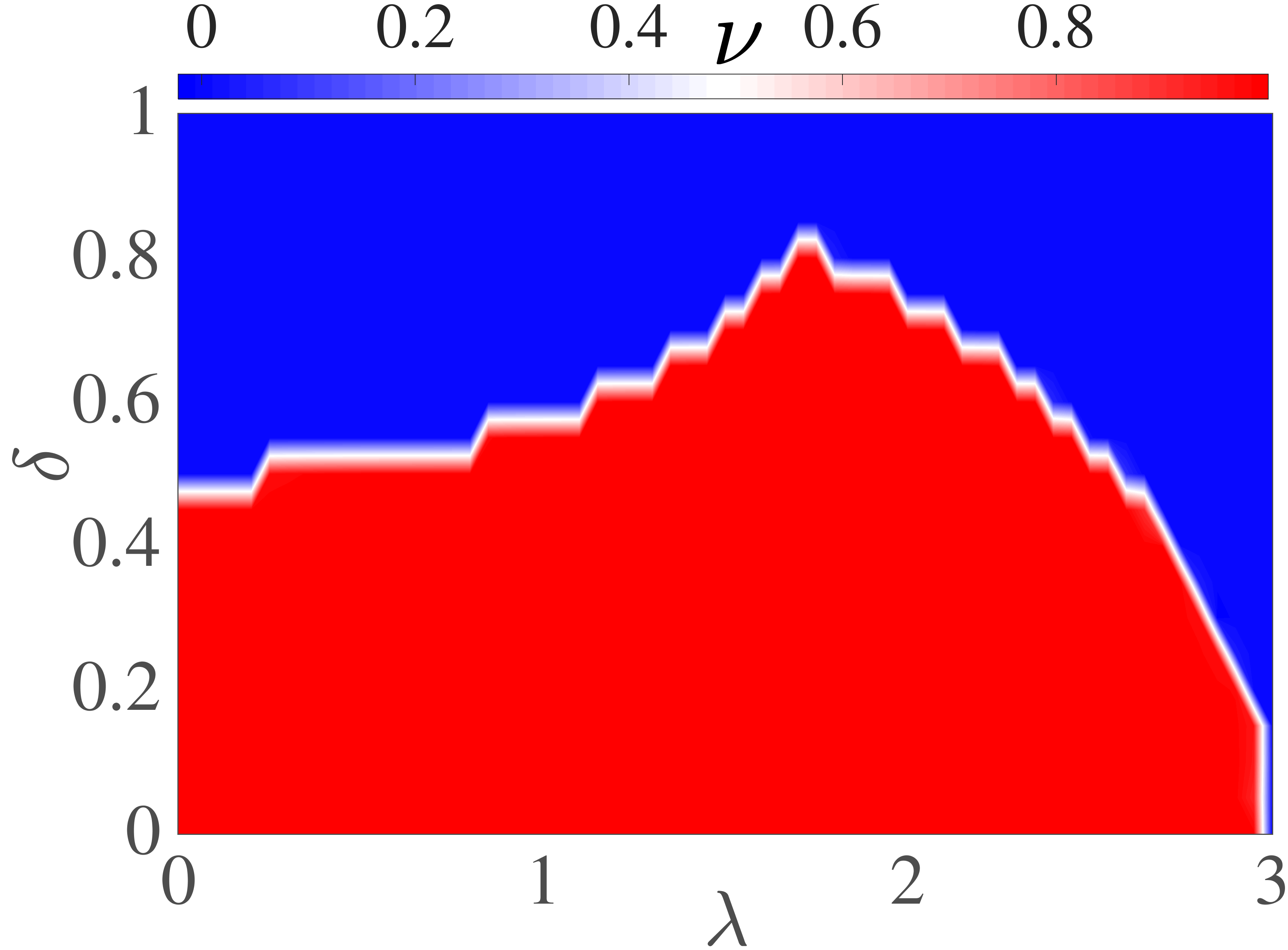}
\hfill}
\caption{The real-space winding number is plotted as a function of $\lambda$. The length of the chain we consider for the calculation is $M=5168$.}
\label{fig:real_space_winding_num}
\end{figure}

\begin{figure*}[!t]
\centerline{\hfill
\includegraphics[width=0.35\textwidth]{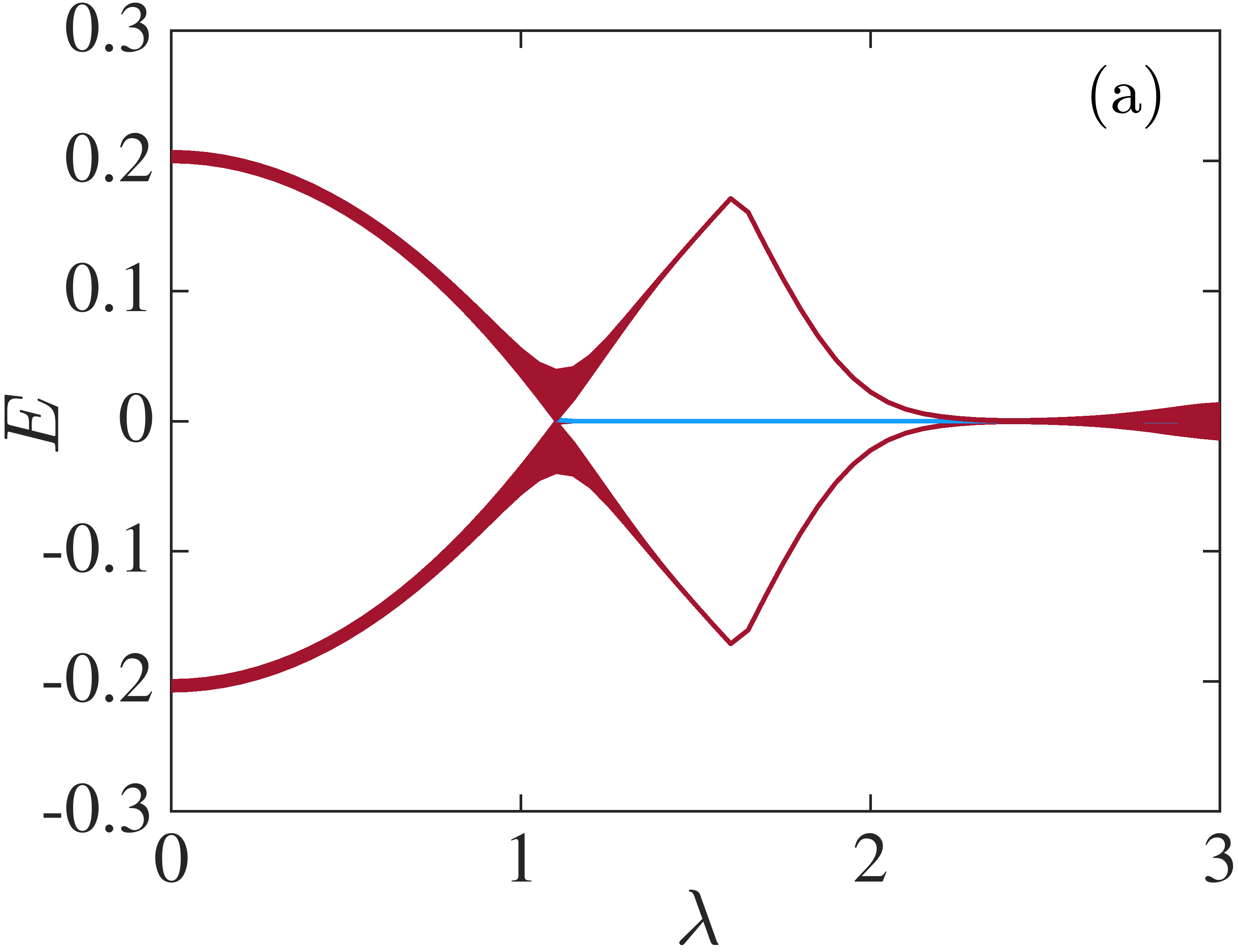}
\hfill
\hfill
\includegraphics[width=0.35\textwidth]{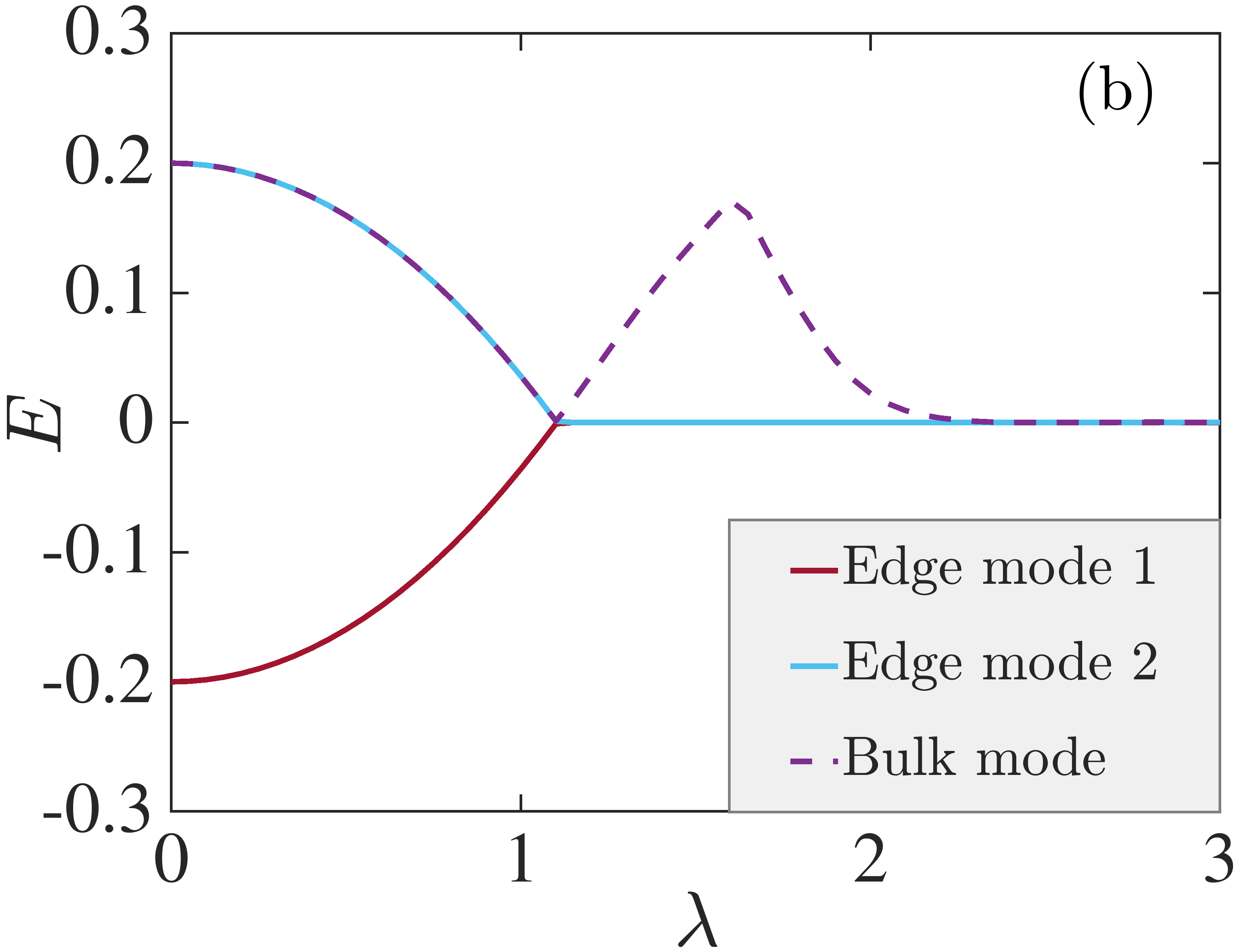}
\hfill
\includegraphics[width=0.35\textwidth]{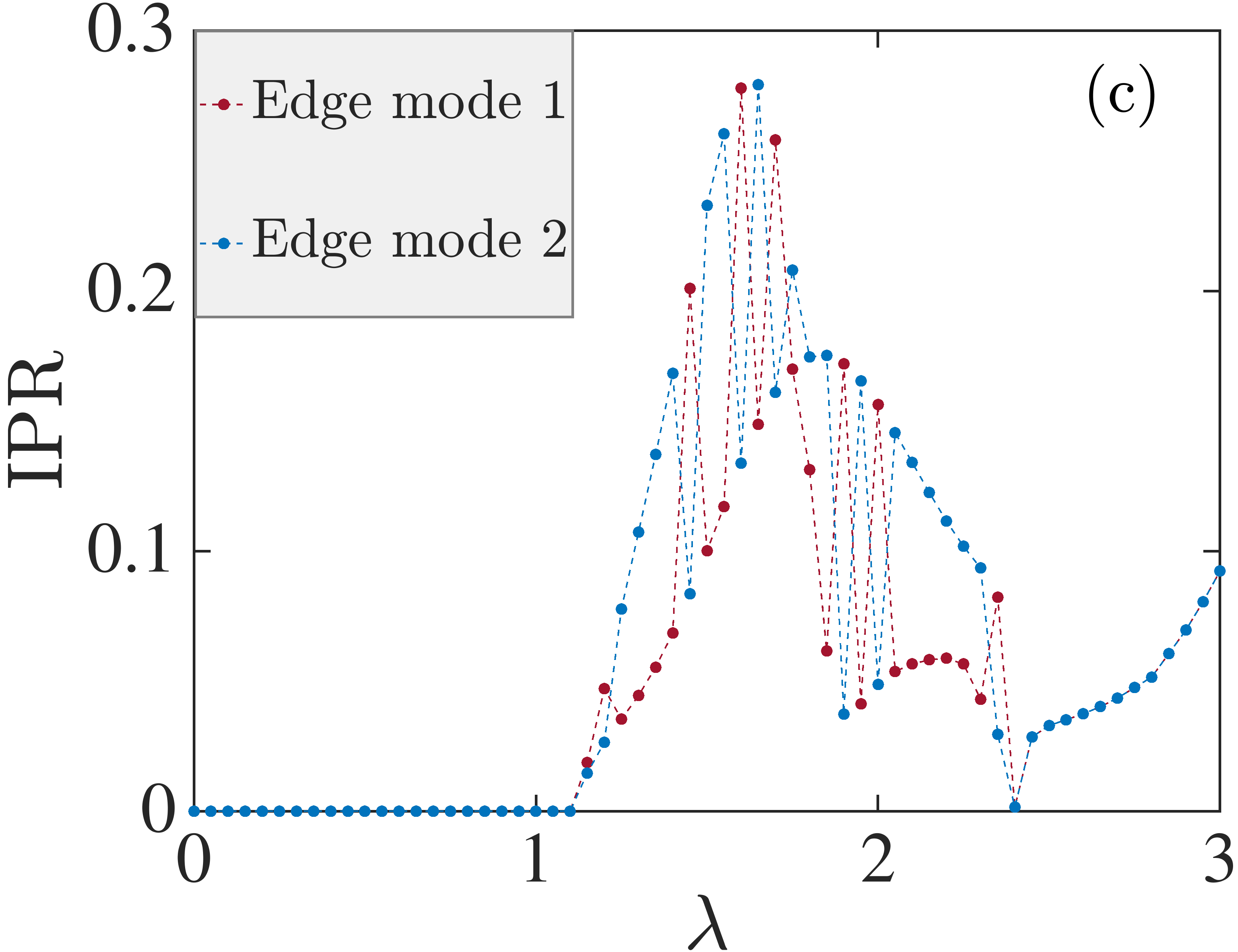}
\hfill}
\caption{Eigenenergies are shown as function of $\lambda$ in (a). Zero-energy edge modes and a bulk mode are plotted as a function of $\lambda$ in (b). IPR value corresponding to the zero-energy edge modes as function of $\lambda$ is shown in (c). The length of the chain we consider for the calculation is $M=8362$. }
\label{fig:bulk_boundary}
\end{figure*}

The nature of an extended state is to spread over the entire lattice, while the localized states only span over a very few lattice sites. In contrast, a multifractal state is fundamentally different from the above two, implying neither an extended nor a localized behavior. To have a clear visualization of the phase transition, we plot the probability distribution of the eigenstates as a function of the site indices in Fig~\ref{fig:bulk_prob}. We observe that, the probability distribution at $\lambda=1.5$ and $\delta=0.6$ (Fig~\ref{fig:bulk_prob} (a)) spreads uniformly over the entire lattice, hence denoting an extended nature. Further, at $\lambda=1.55$ and $\delta=0.6$ (Fig~\ref{fig:bulk_prob} (b)), we observe a fluctuating nature of the states, which aids us in identifying it as multifractal states. Afterwards, at $\lambda=2.4$ and $\delta=0.6$ (Fig~\ref{fig:bulk_prob} (c))  the state is a multifractal, and finally at $\lambda=2.45$ and $\delta=0.6$ (Fig~\ref{fig:bulk_prob} (d)), the states are highly localized and span over only a few of the lattice sites.


Finally, to have a concrete validation of the extended-critical-localized phase transition, we perform a finite-size scaling analysis of the fractal dimension. In order to do that, we calculate $\langle D_{2} \rangle$ using a narrow band consisting of lower energy states, which is plotted as a function of the system lengths. The system sizes are taken as $L=5168,~3194,~1974,$ and $1220$. The intercept of the linear plot will provide the value of $\langle D_{2} \rangle$ in the thermodynamic limit. In Fig~\ref{fig:D2_scaling}, we show the scaling behavior of $\langle D_{2} \rangle$ with the system sizes corresponding to $\lambda=0.5$ (extended) in Fig~\ref{fig:D2_scaling}(a),  $\lambda=1.7$ (critical) in Fig~\ref{fig:D2_scaling} (b), and $\lambda=3.0$ (localized) in Fig~\ref{fig:D2_scaling} (c) for a dimerization strength $\delta=0.6$. Following this, we also show  $\langle D_{2} \rangle$ as a function of $\lambda$ corresponding to various $\delta$ values in Fig~\ref{fig:D2_avD2_lambda} (a). The results clearly distinguish between the three phases by demonstrating a value $1$ for the extended phase, a fractional value (between $0$ to $1$) for the critical phase, and zero for the localized phase, which we have also inferred earlier from the phase diagram presented in Fig~\ref{fig:D2_phasediag} (b).  
In addition to this, we also study the variation of the average $D_{2}$ ($\overline{D_{2}}$) over the upper half of the energy states as a function of $\lambda$ for $L=8362,~5168,~3194,~1974,$ and $1220$ in Fig~\ref{fig:D2_avD2_lambda} (b) corresponding to $\delta=0.6$. The crossing of the curves at $\lambda_{c}\simeq 1$ implies a phase transition from an extended to a critical phase. However, the critical to the localized phase transition is not clearly captured. The results match with the phase diagram presented in Fig~\ref{fig:eta_phasediag}.

\begin{figure}[!t]
\centerline{\hfill
\includegraphics[width=0.5\textwidth]{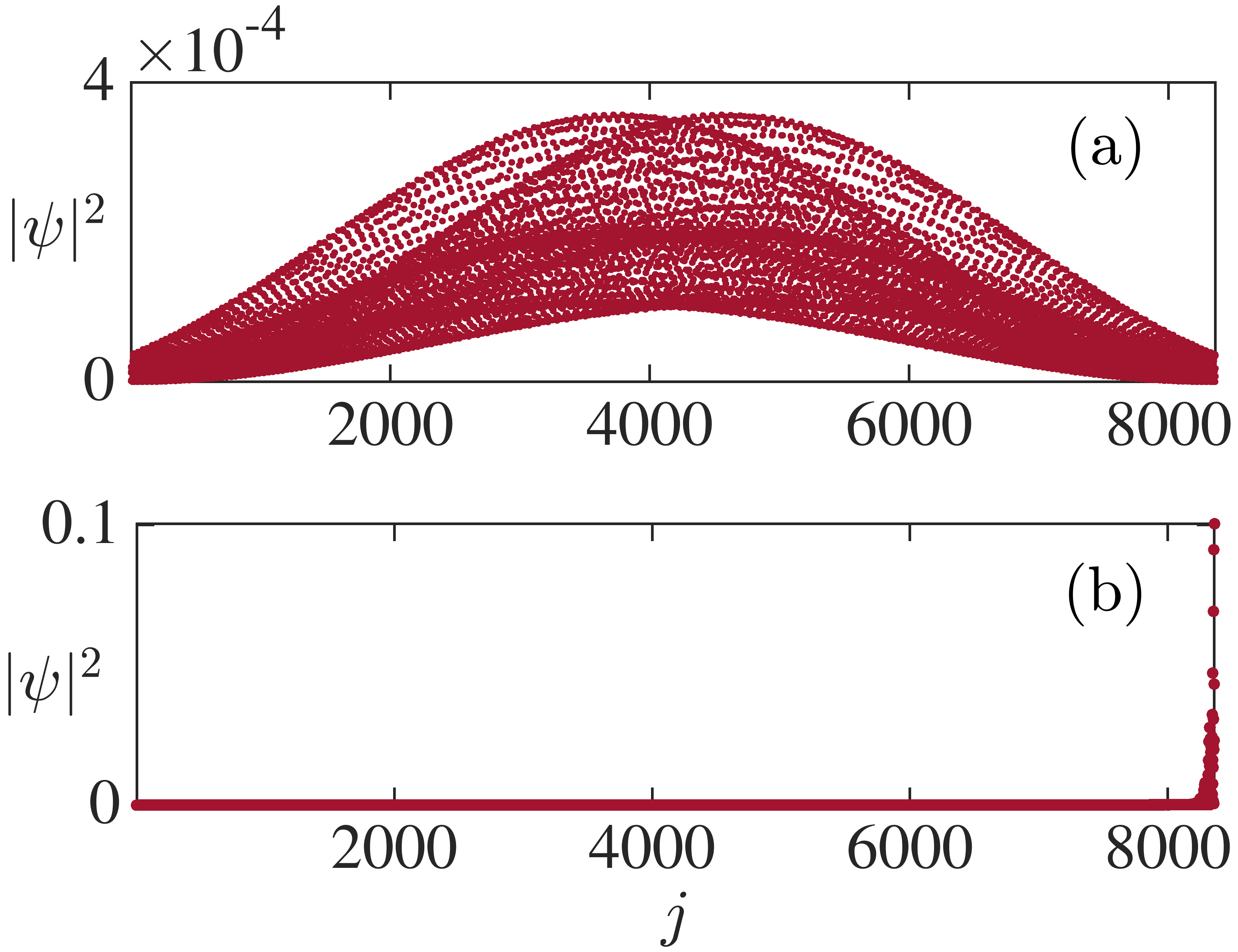}
\hfill}
\caption{The probability distribution of eigenstates (near first transition) as a function of site indices ($j$) are shown corresponding to (a) $\lambda=1.1$, $\delta=0.6$ and (b) $\lambda=1.15$, $\delta=0.6$. The length of the chain we consider for the calculation is $M=8362$. }
\label{fig:first_transition}
\end{figure}

\begin{figure}[!t]
\centerline{\hfill
\includegraphics[width=0.5\textwidth]{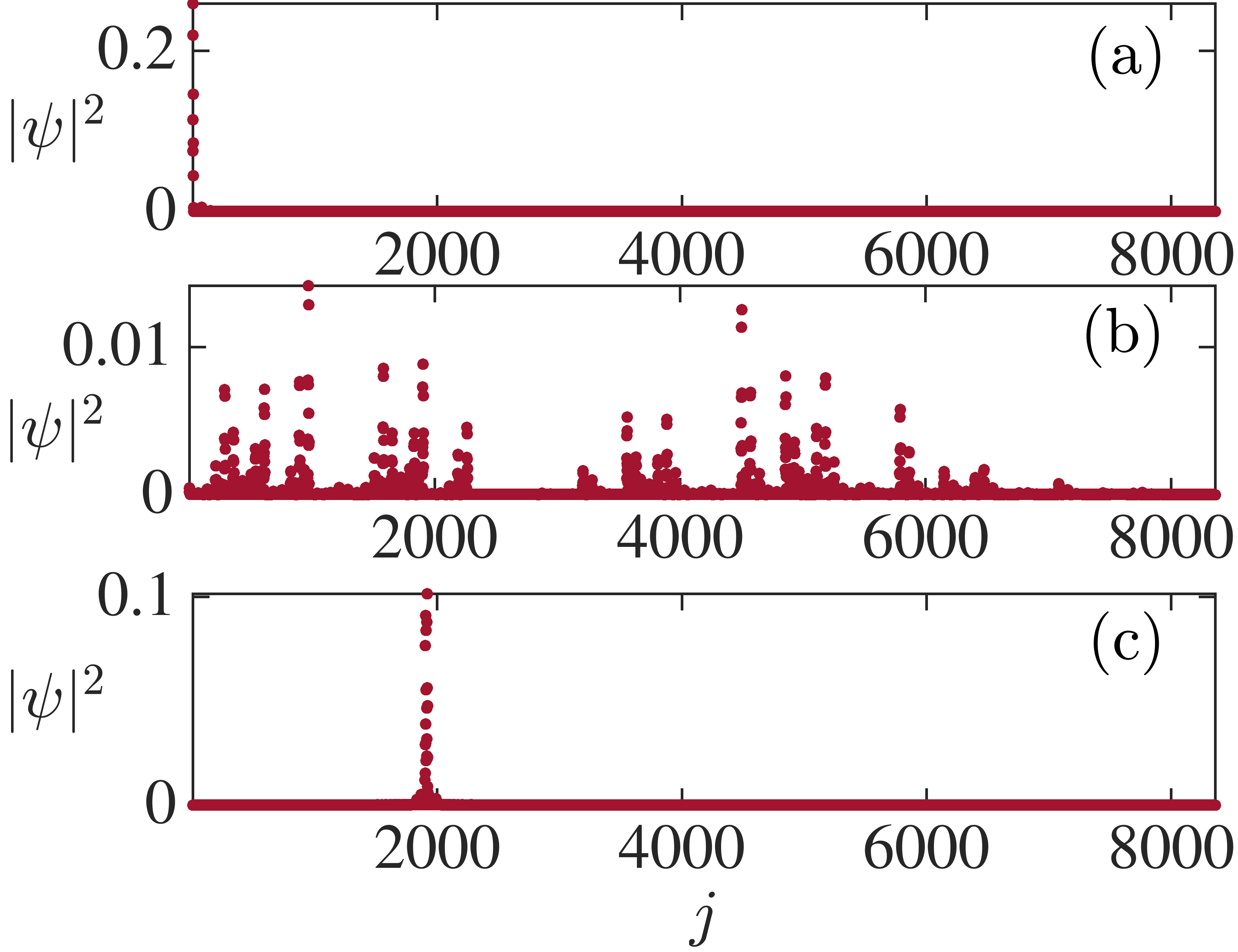}
\hfill}
\caption{The probability distribution of the eigenstates (near second transition) as a function of site indices ($j$) are shown corresponding to (a) $\lambda=2.35$, $\delta=0.6$, (b) $\lambda=2.40$, $\delta=0.6$, and (c) $\lambda=2.45$, $\delta=0.6$. The length of the chain we consider for the calculation is $M=8362$.}
\label{fig:second_transition}
\end{figure}


\begin{figure*}[!t]
\centerline{\hfill
\includegraphics[width=0.33\textwidth]{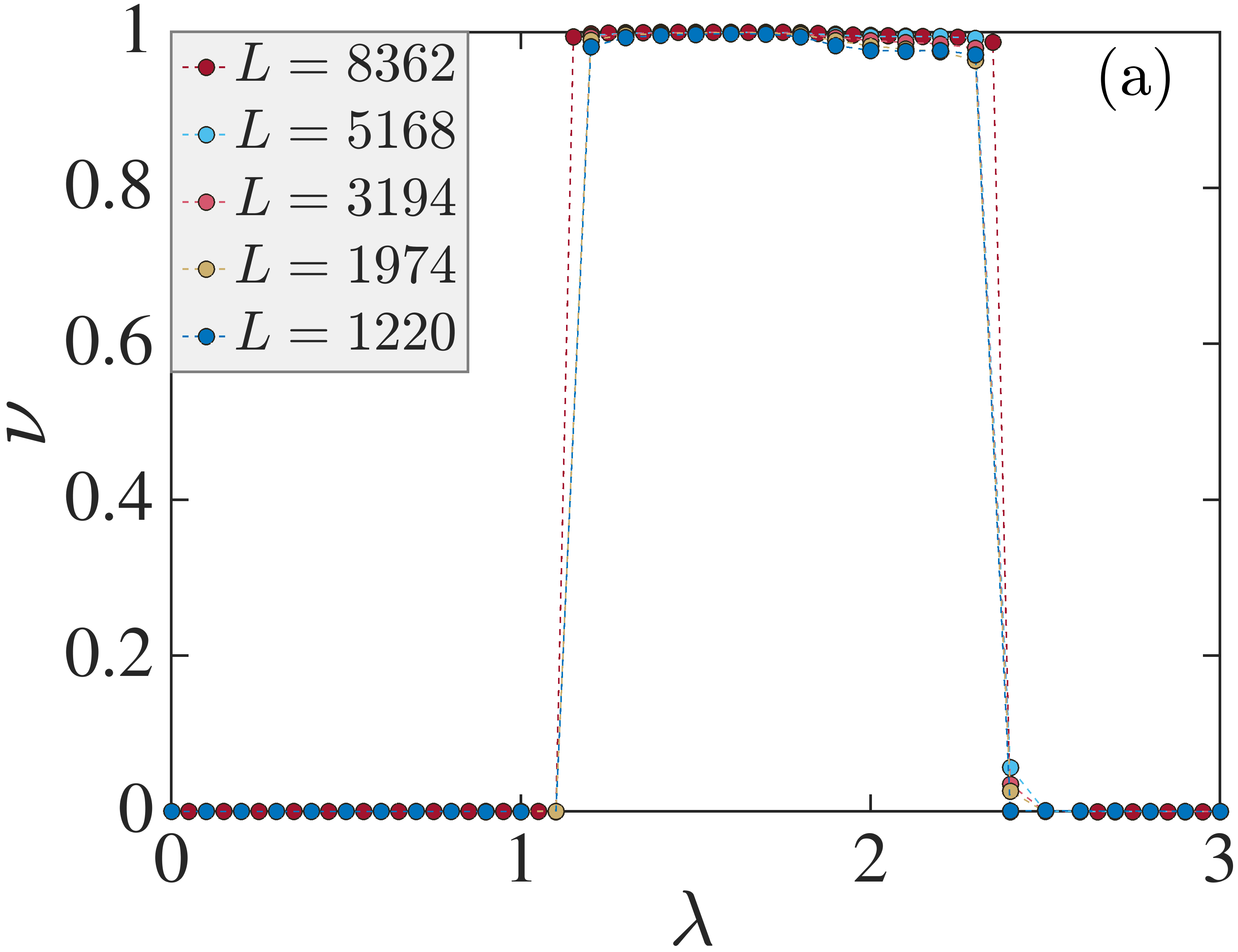}
\hfill
\hfill
\includegraphics[width=0.33\textwidth]{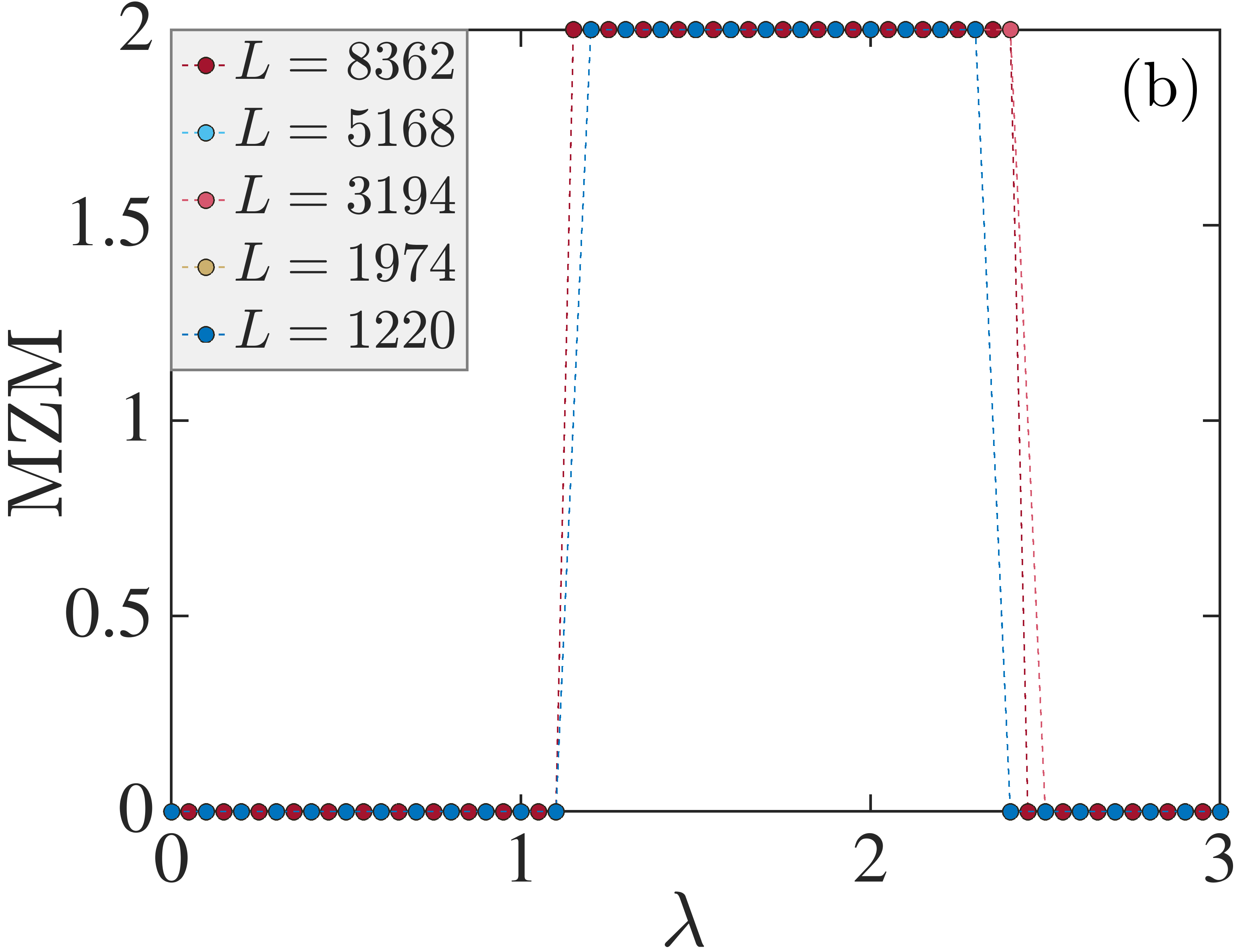}
\hfill
\hfill
\includegraphics[width=0.33\textwidth]{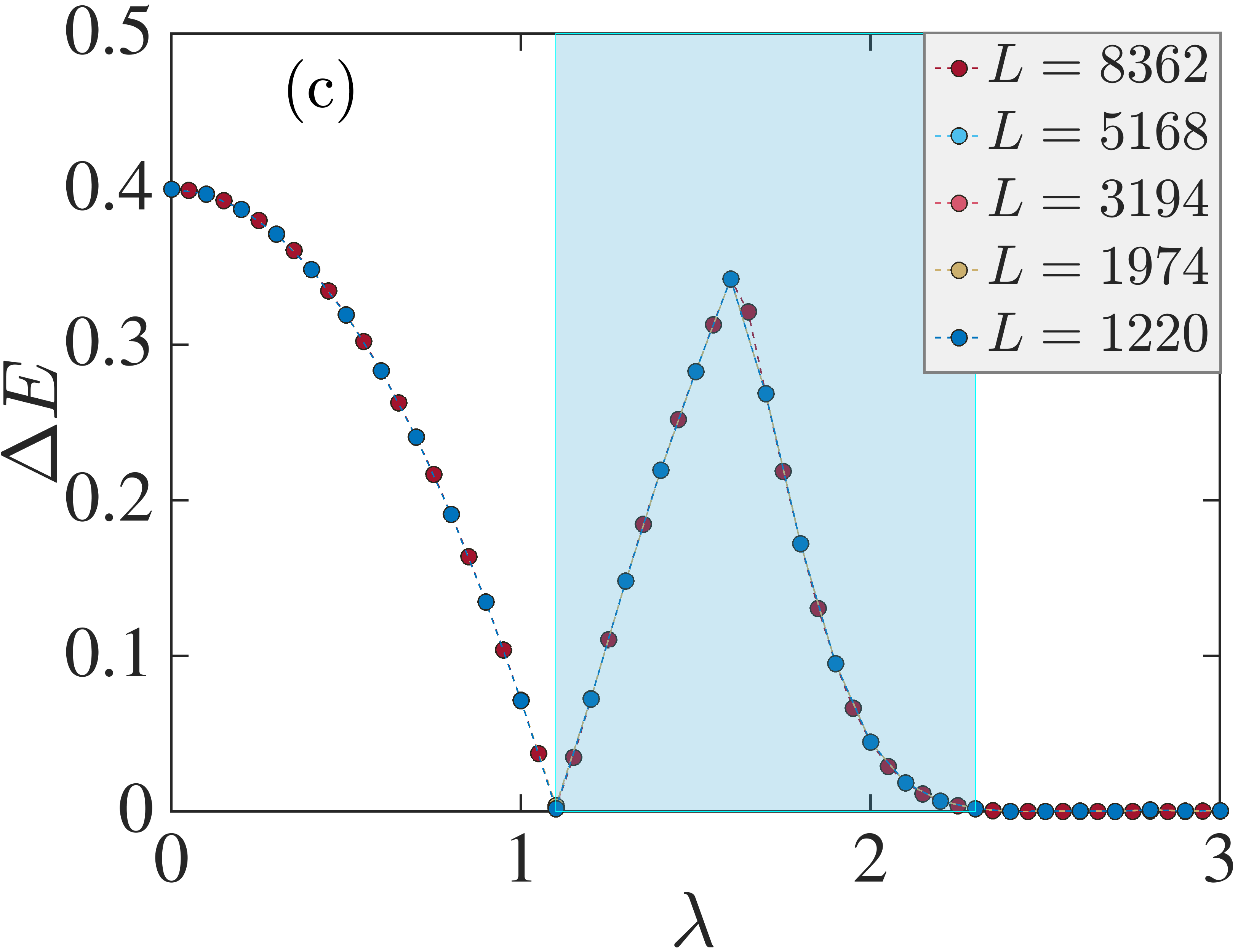}
\hfill}
\caption{The real-space winding number ($\nu$) as a function of $\lambda$ is shown in (a). The number of Majorana zero modes (MZMs) as a function of $\lambda$ is shown in (b). The energy bulk gap $\Delta E$ as a function of $\lambda$ is shown in (c). all the plots are shown for $\delta=0.6$ and various system sizes, mentioned in the figure.  }
\label{fig:finite_size_topology}
\end{figure*}

\subsection{Topological properties of the model}
In this section, we investigate the topological properties of the zero-energy edge modes that emerge in our model. In general, it is known that the topologically non-trivial phase appears to be robust for weak values of potential. However, there is a phase transition from the topological non-trivial to the trivial phase in presence large QP potential. Thus a topological invariant is required to identify them separately. Here, we shall characterize the topological nature by the topological invariant (see below) and the number of the MZMs. Since the potential breaks the translational symmetry, we shall use the real-space winding number as the topological invariant. The real-space winding  number is defined as \cite{PhysRevA.105.063327},
\begin{equation}
\nu=\frac{1}{L^{\prime}} Tr\bigg( \Gamma Q[Q, X]  \bigg)
\label{RSWN}
\end{equation}
where $\Gamma$ and $X$ are the chiral symmetry and the position operator, respectively. The operator, $Q$ can be calculated as,
\begin{equation}
Q=\sum_{j=1}^{L} (|j\rangle \langle j| -|j^{\prime}\rangle \langle j^{\prime}|)
\end{equation} 
where $|j^{\prime}\rangle=\Gamma^{-1} |j\rangle$. $Tr$ represents the trace of the sites with the given length $L^{\prime}=\frac{L}{2}$.

In Fig~\ref{fig:real_space_winding_num}, we show the phase diagram via the real-space winding number $\nu$ as a function of $\delta$ and $\lambda$ using periodic boundary condition . While $\nu$ has a value $1$ corresponding to the topological phase, it is $0$ for the trivial phase. Here, we observe a topologically non-trivial phase up to $\lambda \simeq 3$ corresponding to no dimerization ($\delta=0$). In contrast, the model has a topologically trivial phase at the strong dimerization limit ($\delta=1$). However, a certain region of $\delta$, namely, $0.5<\delta <0.8$ shows an intriguing nature. It is illustrated that, in this region, the system is in a topologically trivial phase for the clean limit ($\lambda=0$). With increasing $\lambda$ and beyond the moderate values, the system enters into a topologically non-trivial regime which spans over a range of $\lambda$. Finally, the model exhibits a transition from the topologically non-trivial to an Anderson localized phase at large QP potential strengths.

The real-space winding number extracts the details of the topological properties using the bulk states of the Hamiltonian. We also need the information on the edge modes to understand the bulk-boundary correspondence. Thus, we show some of the energy eigenvalues around the zero-energy as a function of $\lambda$ corresponding to $\delta=0.6$ in Fig~\ref{fig:bulk_boundary} (a). The result exclusively shows that the bulk gap closes at $\lambda \simeq 1.1$. Later, with increasing $\lambda$, the zero modes persist up to $\lambda \simeq 2.4$. Finally, the two edge modes hybridize and merge with the bulk bands. For a clear visualization, we plot the zero energy edge modes and a single bulk mode in Fig~\ref{fig:bulk_boundary} (b).
Moreover, it is also fascinating to learn about the localization properties of these two edge modes. Thus, we show the IPR value corresponding to both of them as a function of $\lambda$ in Fig~\ref{fig:bulk_boundary} (c). It is observed that the modes are extended in nature with IPR$= 0$ up to $\lambda \simeq 1.1$. With increasing $\lambda$, the states are localized at the two edges of the lattice up to a value such that $\lambda < 2.4$. Beyond $\lambda \geq 2.4$, the edge modes get hybridized with the bulk bands, and a complete Anderson localization transition occurs.

We shall complement these results with the probability distribution of the energy eigenstates corresponding to the zero modes. In Fig~\ref{fig:first_transition}, we plot the  probability distribution of the edge states as a function of the first gap closing point. While in Fig~\ref{fig:first_transition} (a), the eigenstates are distributed uniformly throughout the lattice, Fig~\ref{fig:first_transition}(b) shows that localization occurs at the edges of the lattice, indicating a topological behavior. Further, in Fig~\ref{fig:second_transition}, we plot the probability distributions of three of the eigenstates corresponding to three closely values of second transition point of $\lambda$, namely, $\lambda=2.35,~2.4,$ and $2.45$. Following the previous analysis, the distribution shows that the eigenstates are located at the edges in Fig~\ref{fig:second_transition} (a), thereby demonstrating the presence of the zero-energy edge mode. Fig~\ref{fig:second_transition} (b) shows fluctuations that are occurring across the lattice sites, indicating an emergence of a multifractal behavior, and hence implies the presence of a critical point. Finally, in Fig~\ref{fig:second_transition} (c), the eigenstates show localized behavior that spans over a few of the sites in the bulk of the lattice, signalling the emergence of the localized states due to Anderson localization transition.

Finally, we show the finite size analysis to characterize the topological phase transition corresponding to $\delta=0.6$ in Fig~\ref{fig:finite_size_topology}. In this calculation, we chose the system sizes, such as, $L=8362,~5168,~3194,~1974,$ and $1220$, which are shown with different colors. In Fig~\ref{fig:finite_size_topology} (a), we plot the real-space winding number $\nu$ as a function of $\lambda$ for different system sizes. The variation of $\nu$ with $\lambda$ shows an sharp transition from $0$ to $1$ (beyond $\lambda=1$), indicating a phase transition from a topologically trivial to non-trivial (Topological Anderson) at a particular value $\lambda_{1}$. The phase appears to persist up to a value $\lambda_{2}$ for all system sizes. Later, a second transition occurs, from a topologically non-trivial (Topological Anderson) to an Anderson localized phase. The same behavior is also obtained from the plot of the counts of the number of zero energy edge modes (Majorana zero modes (MZM)) as a function of $\lambda$ in Fig~\ref{fig:finite_size_topology} (b). The information exactly matches the results from the real-space winding number calculations. We also calculate the bulk gap $\Delta E=E_{2}-E_{1}$ in Fig~\ref{fig:finite_size_topology} (c) where $\Delta E=0$ represents the gap closing points. Here we show the presence of the topologically non-trivial phase via a shaded region that in blue color.

\section{Conclusion}
In this work, we have studied the localization and topological properties of a one-dimensional dimerized Kitaev chain in the presence of an onsite QP potential. The localization properties demonstrate phase transitions from an extended to  a critical and hence to a localized phases due to the competition between the dimerization strength and the QP potential strength. One of the prime observations is the existence of the critical phase comprising of two different mobility edges separating the extended-localized and critical-localized phases. Hence a broad region of the critical states are results in a multifractal phase. Additionally, the topological properties of the model computed via the winding number in real-space and the number of Majorana zero modes existence in the system, have shown that a moderate value of the QP potential can induce a topologically trivial to a non-trivial (Topological Anderson) phase transition beyond a certain critical dimerization strength. Beyond this, a non-trivial phase will undergo another transition to the Anderson localized phase at large values of the QP potential.

\bibliography{Dimer_Kitaev_AA}
\end{document}